%% file: main.tex
\documentclass[iop]{emulateapj}
\usepackage{amsmath}
\usepackage{inputenc}
\usepackage{array}   
\usepackage{hhline}
\usepackage{color}
\usepackage[sort&compress]{natbib}

\newcommand{\mbfF}{\mathbf{F}}
\newcommand{\mbfV}{\mathbf{V}}
\newcommand{\mbfu}{\mathbf{u}}

\newcommand{\mbfk}{\mathbf{k}}
\newcommand{\mbfx}{\mathbf{x}}
\newcommand{\mbfq}{\mathbf{q}}

\newcommand{\mbfnabla}{\mathbf{\nabla}}

\newcommand{\kle}{k\lambda_e}
\newcommand{\kli}{k\lambda_i}
\newcommand{\xinu}{\xi_{\nu}}
\newcommand{\xice}{\xi_{ce}}
\newcommand{\xici}{\xi_{ci}}
\newcommand{\xiei}{\xi_{ei}}
\newcommand{\eps}{\epsilon}
\newcommand{\tilu}{\tilde u}
\newcommand{\tiln}{\tilde n}
\newcommand{\tilte}{\tilde T_e}
\newcommand{\tilti}{\tilde T_i}
\newcommand{\tildeT}{\tilde T}
\newcommand{\tildedeltaT}{\delta\tilde T}

\newcommand{\li}{\lambda_i}
\newcommand{\mbfpi}{\mathbf{\pi}}
\usepackage{tabularx}
\usepackage{graphicx}
\usepackage[breaklinks,colorlinks,citecolor=blue,linkcolor=magenta]{hyperref}

\shorttitle{Acoustic Disturbances in Galaxy Clusters}
\shortauthors{Zweibel et al.}

\begin{document}

\title{Acoustic Disturbances in Galaxy Clusters}

\author{Ellen G. Zweibel\altaffilmark{1,2}}
\author{Vladimir V. Mirnov\altaffilmark{2}}
\author{Mateusz Ruszkowski\altaffilmark{3}}
\author{Christopher S. Reynolds\altaffilmark{4,5}}
\author{H.-Y. Karen Yang\altaffilmark{4,6}}
\author{Andrew C. Fabian\altaffilmark{5}}
\altaffiltext{1}{Department of Astronomy, U Wisconsin-Madison, 475 N Charter St., Madison,
                 WI 53706, U.S.A.}
\altaffiltext{2}{Department of Physics, U. Wisconsin-Madison, 1150 University Ave., Madison, WI 53706 U.S.A.}
\altaffiltext{3}{Department of Astronomy, University of Michigan, Ann Arbor, MI 48109 U.S.A. }
\altaffiltext{4}{Department of Astronomy, University of Maryland, College Park, MD 20742 U.S.A.}
\altaffiltext{5}{Institute of Astronomy, University of Cambridge, Cambridge CB3 OHA, U.K.}
\altaffiltext{6}{Einstein Fellow}

\begin{abstract}
Galaxy cluster cores are pervaded by hot gas which radiates at far too high a rate to maintain any semblance of a steady state; this is referred to as the cooling flow problem. Of the many heating mechanisms that have been proposed to balance radiative cooling, one of the most attractive is dissipation of acoustic waves generated by Active Galactic Nuclei (AGN). \cite{Fabian2005} showed that if the waves are nearly adiabatic, wave damping due to heat conduction and viscosity must be well below standard Coulomb rates in order to allow the waves to propagate throughout the core. Because of the importance of this result, we have revisited wave dissipation under galaxy cluster
conditions in a way that accounts for the self limiting nature of dissipation by electron thermal conduction, allows the electron and ion temperature perturbations in the waves to evolve separately, and estimates
kinetic effects by comparing to a semi-collisionless theory. While these effects considerably enlarge the toolkit for analyzing observations of wavelike structures and developing a quantitative theory for wave heating, the drastic
reduction of transport coefficients proposed in \cite{Fabian2005} remains the
most viable path to acoustic wave heating of galaxy cluster cores.
\end{abstract}


\section{Introduction}\label{s:introduction}

Unopposed radiative cooling in  the intracluster medium (ICM) leads to catastrophic mass accretion rates of up to a thousand solar masses per year \citep{FabianNulsen1977}. This constitutes the classic cooling flow problem. Accretion rates predicted by this model are much larger than those inferred from X-ray observations of clusters (e.g., \citet{Peterson2003}). Moreover, the central temperature does not fall below $\sim$keV and the star formation rates are significantly lower than predicted by the cooling flow model (e.g., \citet{Hoffer2012, Donahue2015}).\\
\indent
Several ICM heating mechanisms have been proposed to offset radiative cooling losses in order to solve the cooling flow problem. Below we list various heating mechanisms considered in the literature, discuss their limitations, and identify a few of the most promising ones. New developments in the field suggest that alternative heating modes, that incorporate plasma effects that go beyond pure hydrodynamics, may be needed to better explain ICM heating. One such mechanism is the dissipation of acoustic waves excited by the supermassive black holes in cluster centers, which is the main focus of this paper.  

\subsection{ICM heating mechanisms}
\subsubsection{Thermal conduction}
Thermal conduction from the hot outer ICM regions to the centers of cluster cool cores was considered, e.g., by \citet{ZakamskaNarayan2003}. This mechanism lacks a feedback loop that could maintain cluster atmospheres in globally stable states -- the models either eventually lead to catastrophic cooling or, if conduction is strong, to isothermality of the ICM (e.g., \citet{BertschingerMeiksin1986}), though the thermal runaway may occur on rather long timescales (e.g., \citet{KimNarayan2003}). Because of the self-limiting nature of heat conduction, \citet{YangReynolds2016a} showed that conductive heating is unlikely to be the dominating heating mechanism even for unsuppressed parallel conductivity. Recent results by \citet{RobergClark2016,RobergClark2017,Komarov2017,Fang2018} suggest that the conductive flux may not be linearly proportional to the temperature gradient and that conduction could be severely suppressed compared to the Braginskii level. Thus, conduction on its own does not appear to be a viable solution to the cooling flow problem.

\subsubsection{Dynamical friction and turbulent diffusion}
Dynamical friction acting on galaxies has been considered by \citet{ElZant2004} and \citet{Kim2005}. While this mechanism can be self-regulating because the heating occurs for supersonically moving galaxies,
the models are not thermally stable. Moreover, the minimum temperatures, $\sim T_{\rm vir}$, predicted by this model are larger than observed. While this mechanism is unlikely to provide a complete solution to the cooling flow problem, the onset of thermal instability can be significantly delayed.\\
\indent
\citet{RuszkowskiOh2011} suggested that turbulent heat diffusion may lead to efficient heating of cool cores by redistributing the energy from outer parts of the cool cores to the center. In their model, turbulence is excited by galaxy motions and is volume filling due to the excitation of large-scale $g$-modes. The efficiency of turbulent diffusion is boosted by thermal conduction that reduces the stabilizing buoyancy forces. While there exists a parameter regime for which catastrophic cooling can be avoided or significantly delayed, this is unlikely to be a general solution to the cooling flow problem, especially if thermal conduction is significantly suppressed.

\subsubsection{Cosmic ray heating} The possibility of heating  cool cores by cosmic rays was studied using analytical approaches  (\citet{LoewensteinZweibelBegelman1991,GuoOh2008, Pfrommer2013,JacobPfrommer2017a,JacobPfrommer2017b}) and MHD simulations \citep{RuszkowskiYangReynolds2017}. The emerging consensus from these studies is that cosmic rays may provide sufficient heating to offset radiative cooling even when only a small amount of pressure support in the ICM comes from cosmic rays. While detailed comparisons to the data remain to be performed, these models do not demonstrably violate constraints from radio and gamma-ray observations. Interestingly, these findings are consistent with suggestions by
\citet{Bambic2017}, who study turbulence driving in magneto-hydrodynamical simulations and argue that turbulence driving is inefficient. They suggest that cosmic rays and sound waves may be necessary to model energy thermalization.

\subsubsection{AGN heating}
By far the most promising models to explain the cooling flow problem involve heating by active galactic nuclei (AGN). This mechanism provides a natural self-regulating feedback loop (e.g., \citet{Reynolds2002,RuszkowskiBegelman2002,Guoetal2008,GRS2012,Li2015,YangReynolds2016b}). While the amount of energy supplied by the AGN suffices to offset radiative cooling in cool cores, it is unclear how the AGN energy is distributed and thermalized in the ICM and to what extent the heating is offset by
bubble driven expansion of the overlying gas \citep{GuoMathews2010} .  Here we distinguish between three forms of coupling: radiative heating, mechanical heating by turbulent, or incoherent motions, and mechanical heating or energy transport by coherent flows. Heating by cosmic rays could be considered a fourth type of AGN heating if the cosmic rays are produced by the AGN. \\

\paragraph{Radiative heating} Radiative heating was studied by a number of authors (e.g., \citet{CiottiOstriker2007, CiottiOstrikerProga2010}), who concluded that both AGN mechanical and radiative feedback are needed to prevent catastrophic cooling flows in elliptical galaxies. Recent results by \citet{Xie2017} suggest that radiative heating could be important in low-luminosity AGN, where the kinetic feedback mode is typically considered. They suggest that Compton temperatures in these objects can be $\sim$20 higher than previously assumed and, consequently, AGN can heat the gas radiatively. 

\indent
\paragraph{Turbulent dissipation} 
\citet{Zhuravleva2014} proposed that dissipation of turbulence could offset radiative cooling inside cool cores. Their approach relies on the conversion of gas density fluctuations to the velocity field. Several simplifying assumptions are made, including isotropic turbulence and  absence of gas density fluctuations associated with dark matter substructure. Such fluctuations could mimic turbulent velocity perturbations. Furthermore, dissipation of heat due to mechanisms not involving turbulence could {\it drive} motions and the dissipation of these motions could be interpreted as turbulent dissipation. While the balance of heating and cooling predicted by this model is very approximate, even perfect balance would not necessarily imply that turbulent dissipation is the dominant heating mechanism, as the cluster can go though phases of overheating \citep{LiRuszkowskiBryan2017}. \\
\indent
Resonant scattering can be used to place further constraints on the turbulent velocity magnitude (e.g., \citet{HitomiCollaboration2016, Ogorzalek2017}). Lines of abundant ions can have optical depths exceeding unity and such lines will be attenuated. Turbulence broadens the lines and thus lowers their optical depth and reduces this suppression effect. However, this suppression can be mimicked by predominantly non-turbulent radial gas velocities \citep{Zhuravleva2011}, thus reducing the need for substantial turbulence and associated  turbulent dissipation. This could occur if the AGN jet activity is accompanied by a wide angle wind originating from the vicinity of the central black hole. \\
\indent
In addition to the above caveats concerning the turbulent dissipation model, there are theoretical arguments suggesting the AGN are not likely to drive enough turbulence in the ICM to offset cooling. \citet{Reynolds2015} isolated the role of incompressible modes ($g$-modes and turbulence) using controlled numerical experiments and demonstrated that the energy transfer from the AGN to the ICM is insufficient to balance cooling. This claim was corroborated by a more realistic treatment of AGN feedback in global hydrodynamical simulations of cool cores by \citet{YangReynolds2016b}, who showed that turbulent dissipation contributes to the heating balance at the level of just a percent. The main heating in their model was due to a combination of shocks and mixing. However, mixing of the thermal bubble gas may be partially inhibited by magnetic fields \citep{Ruszkowski2007} and the bubbles may be predominantly filled with cosmic rays rather than thermal gas \citep{DunnFabian2004,GuoMathews2011,Guo2016}.
The above considerations suggest that alternative heating modes need to be explored to explain the thermalization of the energy injected by the AGN in the ICM. \\

A variant on turbulent heating is given in \citet{Kunzetal2011}. This paper is based on the idea that large scale turbulence in the cluster causes the plasma pressure to become anisotropic with respect to the ambient magnetic field ($p_{\perp}\neq p_{\parallel}$; \citet{SchekochihinCowley2006}). The level of anisotropy is determined by a balance between turbulent driving and collisional relaxation. If it is assumed that the resulting anisotropy is at the critical level for the mirror ($p_{\perp}> p_{\parallel}$) or firehose  ($p_{\perp}< p_{\parallel}$) instability, and that relaxation is due to Coulomb collisions, then the resulting heating rate depends only on the ambient magnetic field strength and plasma temperature, and is argued to be thermally stable.
\indent
\paragraph{Shock dissipation} Shocks driven by AGN have been identified through observations \citep{Randalletal2011} and in simulations \citep{YangReynolds2016b}. They tend to
form and dissipate close to their source, and are thus a strongly centrally concentrated form of AGN heating.
However, the shock heated gas may propagate energy away from the cluster
center through time dependent flows \citep{Guoetal2018} that provide a feedback loop.

There is a close relationship between shock waves and sound waves. Sound waves can steepen into weak
shock waves, although geometrical divergence and dissipation counter this effect. On the other hand,  reflection of shock waves from inhomogeneities can produce sound waves. Given their close coupling, sound waves and shock waves should be discussed in tandem.
\indent
\paragraph{Sound wave dissipation} Sound waves in the ICM were first reported in the Perseus cluster by \citet{Fabian2003}, who suggested that viscous dissipation of the waves could balance cooling. Subsequently, \citet{Forman2005} detected waves in the Virgo cluster. Simulations of ICM heating by viscous dissipation of sound waves were first performed by \citet{RuszkowskiBruggenBegelman2004a,RuszkowskiBruggenBegelman2004b}, who found that the waves can indeed heat the gas efficiently and propagate to large distances despite somewhat overheating the very central cluster regions. However, as mentioned above, such overheating may not be inconsistent with the data \citep{LiRuszkowskiBryan2017}. In the simulations of \citet{RuszkowskiBruggenBegelman2004a,RuszkowskiBruggenBegelman2004b}, dissipation due to thermal conduction was completely suppressed. Using linear/analytic arguments, \citet{Fabian2005} suggested that  suppression of transport, and in particular thermal conduction, is needed to allow the waves to propagate far from the AGN as is observed. This suppression also allows for better spatial redistribution of the wave energy without overheating the ICM. 

Additional arguments in favor of sound wave dissipation comes from recent {\it Hitomi} constraints on the low level of turbulence in the ICM \citep{HitomiCollaboration2016, Fabian2017}. These constraints can be most easily satisfied when sound wave dissipation is invoked because the velocity perturbations associated with sound waves are significantly subsonic. However, \citet{ZuHone2017} show that projection effects could hide faster motions. In their analysis, they simultaneously account for the appearance of the spiral features seen in Perseus and match the line velocity shifts. For different lines of sight, velocities can be larger. The observational results may also be biased toward brighter regions and, consequently, do not constrain the velocities in the lower density gas. Nevertheless, acoustic wave dissipation is a promising mechanism because it is consistent with the {\it Hitomi} data and may account for spatially well-distributed heating. While  it is not universally agreed that sound waves are generated efficiently in AGN outbursts \citep{TangChurazov2017} and much remains to be understood about the frequency and power spectrum of such waves, there is enough observational and theoretical evidence for such waves to warrant close examination.\\


The purpose of this paper is to improve the toolkit for studies of acoustic waves in galaxy cluster cores by 
including physical processes that were omitted from previous work. We solve separate equations for electron and ion temperature perturbations, allowing for the possibility that they differ, and we include ion thermal conduction as well as ion viscosity. We consider the transition in electron behavior from nearly adiabatic to nearly isothermal, and show how this reduces the damping rate. Finally, we consider the transition from collisional to collisionless behavior and compare the predictions of kinetic and fluid theory.

 In \S\ref{s:collisionality} we present basic formulae for Coulomb processes in a hydrogen plasma and evaluate them for densities and temperatures derived for the ICM of the galaxy cluster A2199, the properties of which we will continue to use for numerical examples throughout the paper. In \S\ref{s:dispersionrelation} and its
subsections we given an overview of wave propagation, derive and solve the dispersion relation in various  limits,  compare the results with a kinetic theory that includes
collisions, and evaluate the electron and ion temperature perturbations. In \S\ref{s:attenuation} we evaluate the attenuation in amplitude of a propagating wave due to dissipation, and \S\ref{s:heating} we evaluate the rates of entropy production by the various dissipation mechanisms. In 
\S\ref{s:conclusion} we summarize the results and conclusions.


\input{waves_10302017_Section2}

\input{waves_10302017_Section3}

\input{waves_10302017_Sections45}

\clearpage

\end{document}

%% file: waves_10302017_Section2.tex
\section{The Collisionality of the ICM}\label{s:collisionality}

We take a ``collision" to be a random event that perturbs the trajectory of a particle. In the cases considered here, each collision has a small effect and the ``collision time" is the time it takes for many collisions to give an rms change of order unity. 

The propagation and dissipation of waves
 depend critically on the collisionality  of the medium. Thermal conduction, viscosity, and electron-ion heat exchange all dissipate wave energy and are mediated by collisions. In a collisionless plasma, waves are dissipated when particles absorb wave energy through resonances.

Although the role of interactions between particles and microscale waves is under active study (\citet{Kunzetal2011,RobergClark2016,RobergClark2017}), in this paper we derive most numerical estimates from Coulomb collisions\footnote{ \citet{Kunzetal2011} assume the collision frequency is the Coulomb frequency, which implies that the turbulent strain adjusts. \citet{Wieneretal2017} assume the collision frequency adjusts while the strain is externally imposed. This leads to an alternative expression for the mean free path which exceeds the Coulomb mean free path for typical cluster core parameters, suggesting that anomalous collisions are not required to maintain the pressure anisotropy at a stable value.} and describe departures from the Coulomb rates with adjustable parameters as in F05. 

Formulae for the electron and ion Coulomb collision times
$\tau_e$ and $\tau_i$ (which is related to $\tau_e$ by $\tau_i=\tau_e\sqrt{2M/m}$  for ion and electron masses $M,m$ and $T_{e}=T_{i}$)  are given in \citet{Braginskii1965} (hereafter B65). 
In evaluating these formulae we assume a hydrogen plasma, set the Coulomb logarithm $\Lambda=37$, 
express  temperature in units of 10$^7$K, and use $n$ (in cm$^{-3}$) to denote either $n_e$ or $n_i$, resulting in
%
%
%
\begin{eqnarray}\label{tei}
\left(\tau_e,\tau_i\right)=\left(2.44\times 10^8,1.48\times 10^{10}\right)\frac{T_7^{3/2}}{n}{\rm{s}}.
\end{eqnarray}
 It is also useful to have the thermal velocities
\begin{eqnarray}\label{vei}
 (v_e,v_i) \equiv (\sqrt{k_BT/m},\sqrt{k_BT/M} )\\=(1.23\times 10^9, 2.87\times 10^7)T_7^{1/2} {\rm{cm s}^{-1}}
 \end{eqnarray}
 and mean free
paths $\lambda_{e,i}\equiv v_{e,i}\tau_{e,i}$, which are almost the same for the two species
\begin{equation}\label{lei}
\left(\lambda_e,\lambda_i\right)=\left(3.00,4.25\right)\times 10^{17}\frac{T_7^{2}}{n}{\rm{cm}}.  
\end{equation}
In order to describe thermal conductivity and viscosity we introduce diffusivities $D_{e,i}\equiv \lambda_{e,i}^2/\tau_{e,i}$
\begin{equation}\label{Dei}
D_{e,i}=\left(3.69\times 10^{26},1.22\times 10^{25}\right)\frac{T_7^{5/2}}{n}{\rm{cm}}^2{\rm{s}}^{-1}.
\end{equation}

We use density and temperature profiles for the cluster A2199 in numerical examples. From \citet{Johnstone2002}
$(T_7,n)=(5.0r_2^{0.3},6.0\times 10^{-3}r_2^{-0.75})$, where $r_2$ is radius in units of
100 kpc; these formulae hold for $0.05 <r_2 < 2.0$.  For A2199, eqns. (\ref{tei}) and (\ref{lei}) give in Myr, kpc, and (kpc)$^2$/Myr respectively
\begin{eqnarray}\label{a2199taulambda}
\log{\tau_e}=-1.84+1.2\log{r_2},\\
\log{\tau_i}=-0.057+1.2\log{r_2},\\
\log{\lambda_e}=-0.392+1.35\log{r_2},\\
\log{\lambda_i}=-0.240 + 1.35\log{r_2},\\
\log{D_e}=1.06+1.5\log{r_2},\\
\log{D_i}= -0.420+1.5\log{r_2}.
\end{eqnarray}
\begin{figure}[h!]
\begin{center}
\includegraphics[height=45mm]{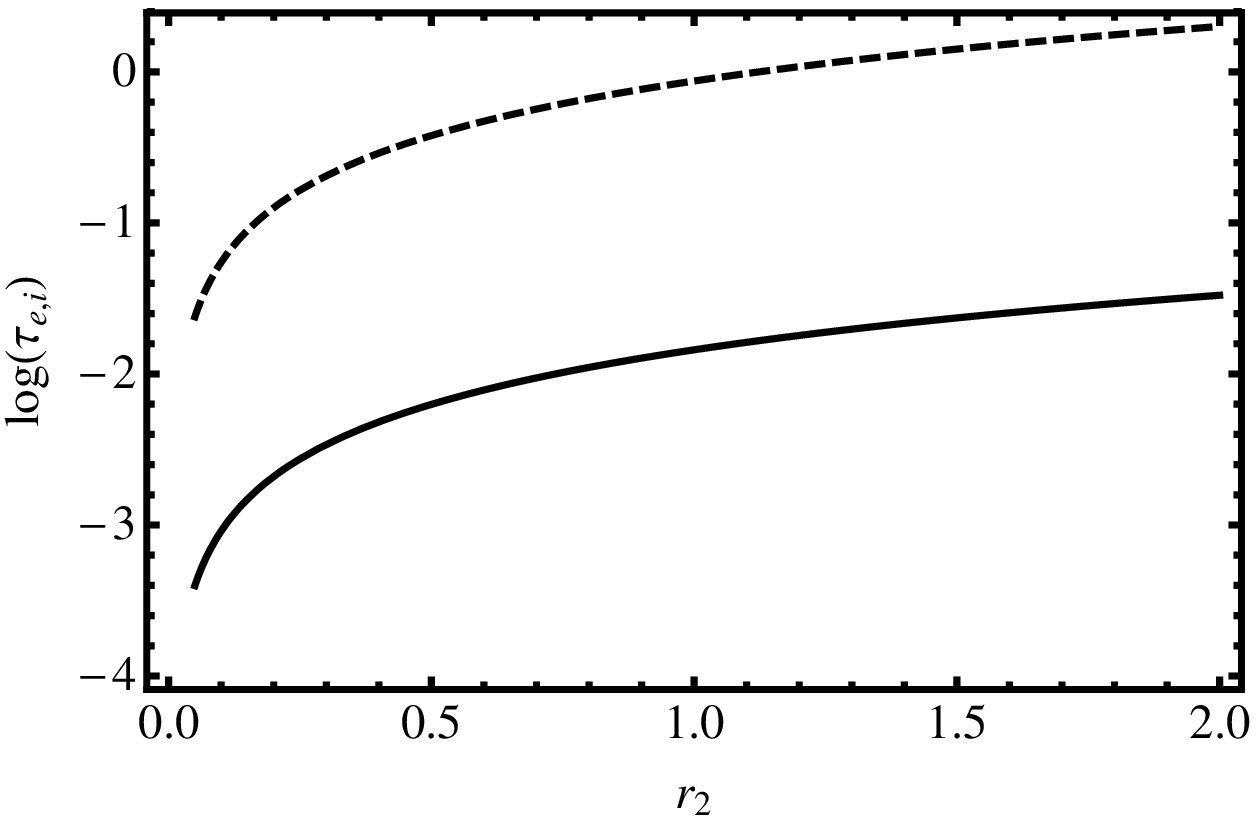}
\includegraphics[height=45mm]{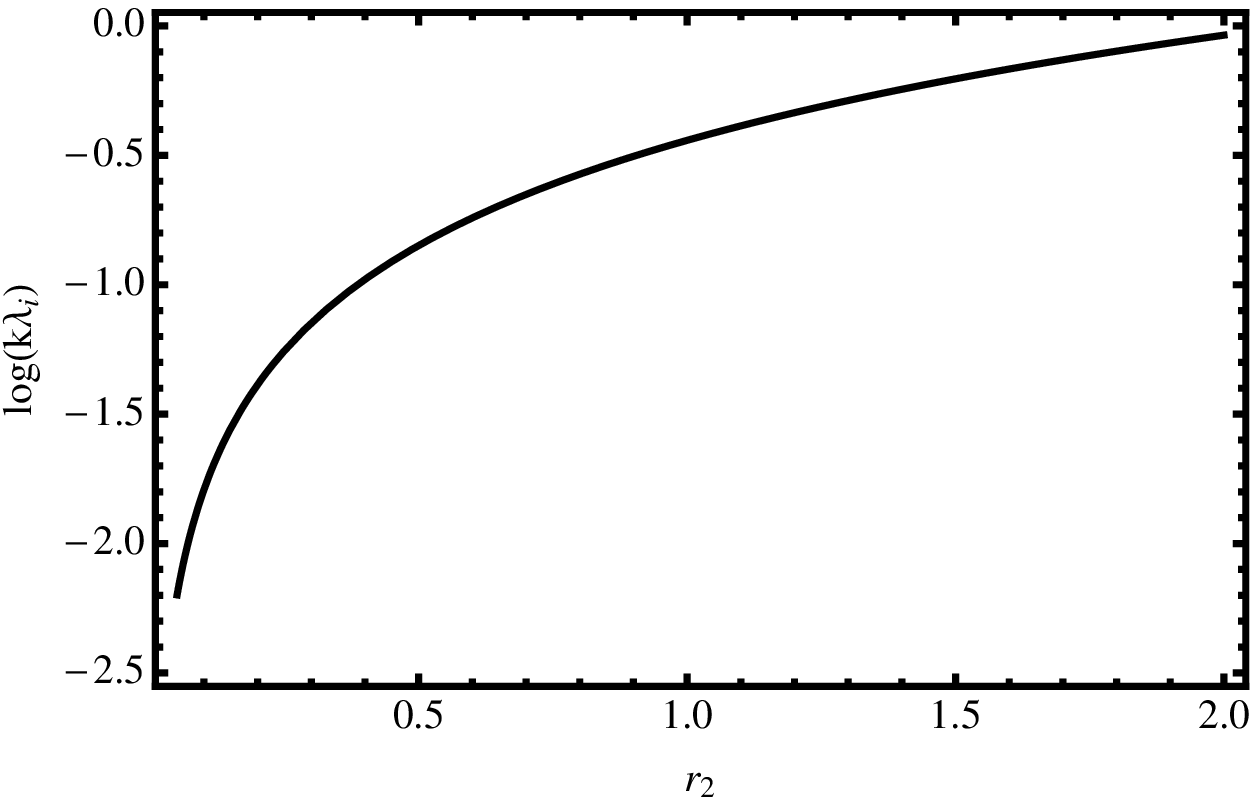}
\caption{Top: Log$_{10}$ of the electron (solid) and ion (dashed) collision times in Myr as functions of radial position in \textcolor{red}{100} kpc for the profiles of $n$ and $T$ measured in A2199 by \citet{Johnstone2002}. Bottom: Log$_{10}$ of the collisionality parameter $k\lambda_i$ vs radial position in kpc, assuming a wavelength of 10kpc, for A2199. The damping rate of such a  wave can be estimated
accurately  with
fluid theory  within about 100 kpc of the cluster, but should be calculated from kinetic theory beyond (see Figure \ref{figure6aab}) .}
\label{figure1aab}
\end{center}  
\end{figure}
The electron and ion collision times and the dimensionless parameter $k\lambda_i$ for a wave with wavenumber $k=2\pi/(10 {\rm{kpc}})$  for A2199 are plotted in Figure \ref{figure1aab}. 

%% file: waves_10302017_Section3.tex
\section{Basic Equations and Dispersion Relation}\label{s:dispersionrelation}
\subsection{Formulation and Estimates}\label{ss:overview}

We consider longitudinal ($\mbfk\times\mbfu\equiv 0$) electrostatic waves of sufficiently low frequency that  electron inertia can be neglected. With these assumptions the wave electron pressure gradient force is almost exactly balanced by the force from the wave  electric
field. Ion motion  is driven by the fluctuating ion pressure gradient and electric field, which due to electron force balance is equivalent to driving by the fluctuating electron pressure
gradient. In these low frequency waves, the electron
 and ion densities are essentially the same (quasi-neutrality). This is the standard propagation regime for ion-acoustic waves in both the fluid and kinetic descriptions, and also holds 
for the thermal and
relaxation waves discussed in \S\ref{ss:fluid}. However, although the electrons and ions are tightly coupled dynamically, they are only coupled thermally through collisions,
and we will see that in general their temperature perturbations are different. 

In a stratified medium with density scale height $H$, the effect of gravity on sound wave propagation appears through the acoustic
cutoff frequency $\omega_{ac}\sim c_s/(2H)$,  which sets a lower
limit on the frequency of a propagating acoustic wave. For the power law density profiles considered here, stratification effects are of order $(kr)^{-1}\ll 1$, and we neglect them. Due to the decrease of $\omega_{ac}$ with $r$, outward propagating waves will not be trapped in an acoustic cavity, but waves generated at large
$r$ might be reflected by the acoustic cutoff barrier as they propagate inward.

We also neglect forces due to magnetic fields. This is strictly accurate only for waves propagating parallel to the background magnetic field, but should be a reasonable approximation if the magnetic field is weak, as is thought to be the case in galaxy clusters. However, even a weak field can drastically affect  plasma transport processes, and our parameterized modification of the transport coefficients
is intended to account for  magnetic geometry as well as
 anomalous collisional processes caused by small scale electromagnetic fluctuations.  Because the cosmic ray pressure in galaxy clusters is also thought to be weak, e.g. \citet{Aleksicetal2012}, we also neglect the thermal and dynamical effects of
cosmic rays.  Finally, we neglect perturbations to the heating and radiative cooling rates because their timescales are long compared to the wave period, because the heating mechanism is unknown, and because whether the cluster gas is in thermal equilibrium at all is uncertain. We briefly discuss the possible effects of magnetic fields, cosmic rays
and thermal damping/instability in \S\ref{s:conclusion}.

As for our estimates of collisional and thermal parameters,  we assume a hydrogen plasma of uniform particle density $n_e=n_i=n$ and temperature $T_e=T_i=T$  and denote the electron and proton masses by $m$ and $M$, respectively.  We introduce $\eps\equiv m/M$, which we will treat as a small parameter.  With this notation, $\tau_e/\tau_i$, $v_i/v_e$, and $D_i/D_e$ are all of order $\eps^{1/2}$. 

 We expect the wave frequencies and wavenumbers of acoustic waves to be related by $\omega\sim kv_i$, and the characteristic timescales
associated with electron thermal conduction, ion thermal conduction, and ion viscosity to be of order $(k^2D_e)^{-1}$, $(k^2D_i)^{-1}$, $(k^2D_i)^{-1}$. These diffusive processes should be important if their timescales are less than the wave period, or $\omega\tau_i>\eps^{1/2}$ for electron thermal conduction and $\omega\tau_i > 1$ for ion thermal
conduction or viscosity.
But because $\omega\tau_i\sim 1$ is roughly equivalent to $k\lambda_i\sim 1$, kinetic effects, viscosity, and heat conduction for ions all become important at similar wavelengths. In \S\ref{ss:kinetic} we show that the ion damping predicted by fluid theory is somewhat larger than that predicted by kinetic theory and is likely an overestimate.  

\subsection{Fluid Theory}\label{ss:fluid}

Following the physical picture described in \S\ref{ss:overview},
we represent the plasma by single momentum and continuity equations, but separate energy equations for the electrons and ions. The evolution of small amplitude perturbations
of a uniform medium is then described by the system of linear equations
\begin{equation}\label{momentum}
nM\frac{\partial\mbfu_1}{\partial t}=-\mbfnabla (P_{e1}+P_{i1}) -\mbfnabla\cdot\mbfpi_1, 
\end{equation}
\begin{equation}\label{state}
P_{e1}+P_{i1}=2n_1T+n(T_{e1}+T_{i1}),
\end{equation}
\begin{equation}\label{continuity}
\frac{\partial n_1}{\partial t}=-n\mbfnabla\cdot\mbfu_1,
\end{equation}
\begin{equation}\label{eenergy}
\frac{3}{2}\frac{\partial T_{e1}}{\partial t}=-T\mbfnabla\cdot\mbfu_1+\chi_e\nabla^2 T_{e1}-
3\frac{m}{M}\frac{(
T_{e1}-T_{i1})}{\tau_e},
\end{equation}
\begin{equation}\label{ienergy}
\frac{3}{2}\frac{\partial T_{i1}}{\partial t}=-T\mbfnabla\cdot\mbfu_1 +\chi_i\nabla^2 T_{i1}+3\frac{m}{M}\frac{(
T_{e1}-T_{i1})}{\tau_e},
\end{equation}
where 
\begin{equation}\label{pi1}
\pi_{1ab}=-\eta_0 \left(\frac{\partial u_{1a}}{\partial x_b}+\frac{\partial u_{1b}}{\partial x_a} - \frac{2}{3}\delta_{ab}\mbfnabla\cdot\mbfu_1\right)\equiv -
\eta_0 W_{1ab}
\end{equation}
is the  stress tensor to first order in wave amplitude,
\begin{equation}\label{eta0}
\eta_0\equiv 0.96nMD_i
\end{equation}
is the ion viscosity,
\begin{equation}\label{chie}
\chi_e\equiv 3.16 D_e
\end{equation}
is the electron thermal conductivity, and
\begin{equation}\label{chii}
\chi_i\equiv 3.90 D_i
\end{equation}
is the ion thermal conductivity. Ion thermal conduction was not included in F05, but it is of the same order as ion viscosity, so we
retain it here. The 
numerical coefficients
multiplying $D_{e,i}$ in eqns. (\ref{pi1}) - (\ref{chii}) are calculated from kinetic theory, and
taken from B65. We have omitted electron viscosity, which is always a minor effect.

We will want to allow for modified transport coefficients, so we introduce parameters $\xice$, $\xici$, $\xinu$, $\xiei$ to multiply $\chi_e$, $\chi_i$, $\eta_0$, and 
the electron - ion
equilibration term in all the equations. The parametric approach is undoubtedly an 
oversimplification: the $\xi$ should be functions that depend on local quantities
such as $n$ and $T$ and possibly also on global properties such as magnetic field
geometry and level of large scale turbulence. In fact, the functional forms of the
$\xi$ may be critical in closing the feedback loop. However, since we have no theory
for the $\xi$, we adopt simple parameterization here.
In principle, the $\xi$ factors could have any magnitude, but we will always assume they suppress transport, i.e. that they lie between $0$ and $1$\footnote{The electron - ion thermal coupling parameter $\xiei$ may be an exception to this; \citet{MarkevitchVikhlinin2007} argued for anomalously fast $T_e$, $T_i$ equilibration in cluster shocks. However it is not clear that the anomalous processes driven in shocks also exist in acoustic waves.}.

We  consider solutions of eqns. 
(\ref{momentum}) - (\ref{ienergy}) which depend on $t$ and $\mbfx$ as $e^{i(kx-\omega t)}$. We will generally follow F05 in treating $\omega$ as known, real , and positive (it represents the frequency at
which the waves are driven) and solving for $k$, which in general is complex; $k=k_r+ik_i$. However in \S\ref{ss:kinetic} we  treat $k$ as real and solve for $\omega$, to facilitate comparison with results from kinetic theory.

We nondimensionalize the problem by normalizing the first order quantities such that $(u_1, n_1, T_{e1}, T_{i1})\rightarrow (u_1/v_i, n_1/n, T_{e1}/T, T_{i1}/T)
\equiv (\tilu, \tiln,\tilte,\tilti)$ and introducing a 
scaled frequency $\Omega\equiv\omega\tau_i\eps^{-1/2}$ and a scaled wavenumber $K\equiv kv_i/\omega$. Ions are collisional ($\omega\tau_i < 1$) for $\Omega < \eps^{-1/2}$
and electrons are collisional for $\Omega < \eps^{-1}$. According to the density and temperature profiles we adopted for A2199,  waves with 10 Myr period span the range $0.64 < \Omega <  54$ in $0.05 < r_2 < 2.0$, so we must consider a large range of propagation conditions.
 Using eqns. (\ref{state}) and (\ref{continuity}) in eqns. (\ref{momentum}),
(\ref{eenergy}), (\ref{ienergy})
and assuming plane wave structure we derive the coupled system
\begin{equation}\label{scaledmomentum}
\left [1-2K^2+1.28 i \xi_{\nu} \epsilon^{1/2} \Omega K^2  \right] \tilde {n} - K^2 (\tilde {T_{e}} + \tilde {T_{i}})=0,
\end{equation}
\begin{equation}\label{scaledeenergy}
\frac{2}{3} \tilde {n} -\left[1+1.49 i \xi_{ce} \Omega K^2 \right] \tilde {T_{e}} -\frac{2.83 i \xi_{ei} }{\Omega} \left(\tilde {T_{e}} -\tilde {T_{i}} \right)=0,
\end{equation}
\begin{equation}\label{scaledienergy}
\frac{2}{3}\tilde {n}  -\left[1+2.60i\xi_{ci}\epsilon^{1/2}\Omega K^2\right]\tilde {T}_{i}-\frac{2.83i\xi_{ei}}{\Omega}\left(\tilde{T}_{i} - \tilde{T}_{e}\right) = 0,
\end{equation}

%
%
%
%
%
%
where we have used the equation of continuity
\begin{equation}\label{scaledcontinuity}
 \tiln -K\tilu = 0
 \end{equation}
 to eliminate $\tilu$.
                                                                                                                                                                                                                                                                                                                                                                                                                                                                                                                                                                                                                                                                                                                                                                                                                                                                                                                                                                                                                                                                                                                                                                                                                                                                                                                                                                                                                                                                                                                                                                                                                                                                                                                                                                                                                                                                                                                                                                                                                                                                                                                                                                                                                                                                                                                                                                                                                                                                                                                                                                                                                                                                                                                                                                                                                                                                                                                                                                                                                                                                                                                                                                                                                                                                                                                                                                                                                                                                                                                                                                                                                                                                                                                                                                                                                                                                                                                                                                                                                                                                                                                                                                                                                                                                                                                                                                                                                                                                                                                                                                                                                                                                                                                                                                                                                                                                                                                                                                                                                                                                                                                                                                                                                                                                                                                                                                                                                                                                                                                                                                                                                                                                                                                                                                                                                                                                                                                                                                                                                                                                                                                                                                                                                                                                                                                                                                                                                                                                                                                                                                                                                                                                                                                                                                                                                                                                 
Equations (\ref{scaledmomentum}) - (\ref{scaledienergy}) describe three distinct linear modes which can be found by
standard linear algebra techniques. 
However, for later purposes (\S\ref{sss:equilibration}) and additional physical insight,  we rewrite eqns. (\ref{scaledeenergy}) and (\ref{scaledienergy}) in terms of the new variables $\tildeT\equiv\tilte+\tilti$, $\tildedeltaT\equiv\tilti-\tilte$ in terms of
which
%
\begin{equation}\label{tilti}
\tilti=\frac{1}{2}\left(\tildeT+\tildedeltaT\right),
\end{equation}
\begin{equation}\label{tilte}
\tilte=\frac{1}{2}\left(\tildeT-\tildedeltaT\right),
\end{equation}
%
Using eqns. (\ref{tilti}) and (\ref{tilte}) in eqns. (\ref{scaledeenergy}) and (\ref{scaledienergy})  leads to a pair of equations for $\tildeT$ and $\tildedeltaT$
\begin{equation}\label{deltaT}
\tildedeltaT-\frac{i\Omega^2 K^2c_{-}}{\Omega+i\Omega^2 K^2c_{+}+5.66i\xi_{ei}}\tildeT=0,
\end{equation}
\begin{equation}\label{tildeT}
\left(1+i\Omega K^2c_{+}\right)\tildeT - i\Omega K^2c_{-}\tildedeltaT = \frac{4}{3}\tiln,
\end{equation}
where the $c_{\pm}\equiv \left( 1.49\xi_{ce}\pm 2.60\xi_{ci}\eps^{1/2}\right)/2$
are proportional to the scaled sum and difference of the electron and ion thermal conductivities. 
Substituting eqn. (\ref{deltaT}) into eqn. (\ref{tildeT}) and using  eqn. (\ref{scaledcontinuity}) leads to expressions for both temperature variables in terms of $\tiln$ 
\begin{equation}\label{tildeT2}
\tildeT=\frac{\Omega+i\Omega^2 K^2 c_{+}+5.66i\xi_{ei}}{\mathcal{D}}\frac{4}{3}\tiln,
\end{equation}
\begin{equation}\label{deltaT2}
\tildedeltaT=\frac{i\Omega^2 K^2c_{-}}{\mathcal{D}}\frac{4}{3}\tiln,
\end{equation}
where
\begin{equation}\label{Denominator}
{\mathcal{D}}\equiv \Omega+2i\Omega^2 K^2c_{+}+\Omega^3 K^4(c_{-}^2-c_{+}^2)+5.66i\xi_{ei}(1+i\Omega K^2c_{+}).
\end{equation}
Substituting eqn. (\ref{deltaT2}) into eqn. (\ref{scaledmomentum}) leads to the dispersion relation
\begin{eqnarray}\label{fulldr}
\left[1-2K^2+1.28i\xinu\eps^{1/2}\Omega K^2\right]{\mathcal{D}}= \nonumber \\
\frac{4}{3}K^2\left(\Omega+i\Omega^2 K^2c_{+}+5.66i\xi_{ei}\right).
\end{eqnarray}

Neglecting all dissipative effects in eqn. (\ref{fulldr}) gives the dispersion relation in the ideal limit
\begin{equation}\label{idealdr}
1-\frac{10}{3}K^2 = 0,
\end{equation}
with solution $K^2\equiv K_0^2 = 3/10$, as expected for acoustic waves in a $\gamma=5/3$ gas with mean particle mass $M/2$. 

When dissipation is included, eqn. (\ref{fulldr})
 is cubic in $K^2$ and describes three distinct wave modes.

%
%
The acoustic mode is of greatest interest here. In the nearly adiabatic limit $\Omega\ll 1$  we can solve for this mode by setting 
 $\tildeT$ equal to its adiabatic value perturbed by electron and ion thermal conduction, $(4/3)\tiln/(1+i\Omega K^2c_{+})$,  in eqn. (\ref{scaledmomentum}) or alternatively keeping only terms proportional to $\Omega^{0}$ and $\Omega$ in eqn. (\ref{fulldr}) and assuming $K$ is order unity.  The resulting approximate dispersion relation is
%
\begin{multline}\label{acousticdr}
1-\frac{K^2}{K_0^2}=\\ -i\Omega K_0^2\left[ 1.28\xinu\eps^{1/2} + (1-2K_0^2)c_{+}\right].
\end{multline}
%
The imaginary terms in square brackets represent,
respectively, ion viscosity and the combined effects of ion and electron thermal conduction. They lead to spatial damping at the rate $K_i$, which is given to first order in
$K_i/K_0$ by
%
%
%
\begin{multline}\label{K1K0}
\frac{K_i}{K_0}= \frac{3}{20}\Omega\left[1.28\xinu\eps^{1/2} + \frac{2}{5}c_{+}\right]\\=\Omega\left(0.0045\xinu+0.045\xice+0.0018\xici\right)
\end{multline}
or, writing the imaginary part of $k$ in terms of the ion mean free path and substituting numerical values for $\eps$ and $K_0$,
\begin{equation}\label{kli_ad}
k_i\li=(\omega\tau_i)^2\left(0.105\xinu+1.03\xice+0.042\xici\right).
\end{equation}
%
Equation (\ref{K1K0}) agrees with eqn.(1) of F05 when written in their notation, except that F05 omitted ion thermal conduction, which increases ion damping by about 40\% if $\xinu=\xici$. 

\begin{figure}[h!]
\begin{center}
\includegraphics[height=45mm]{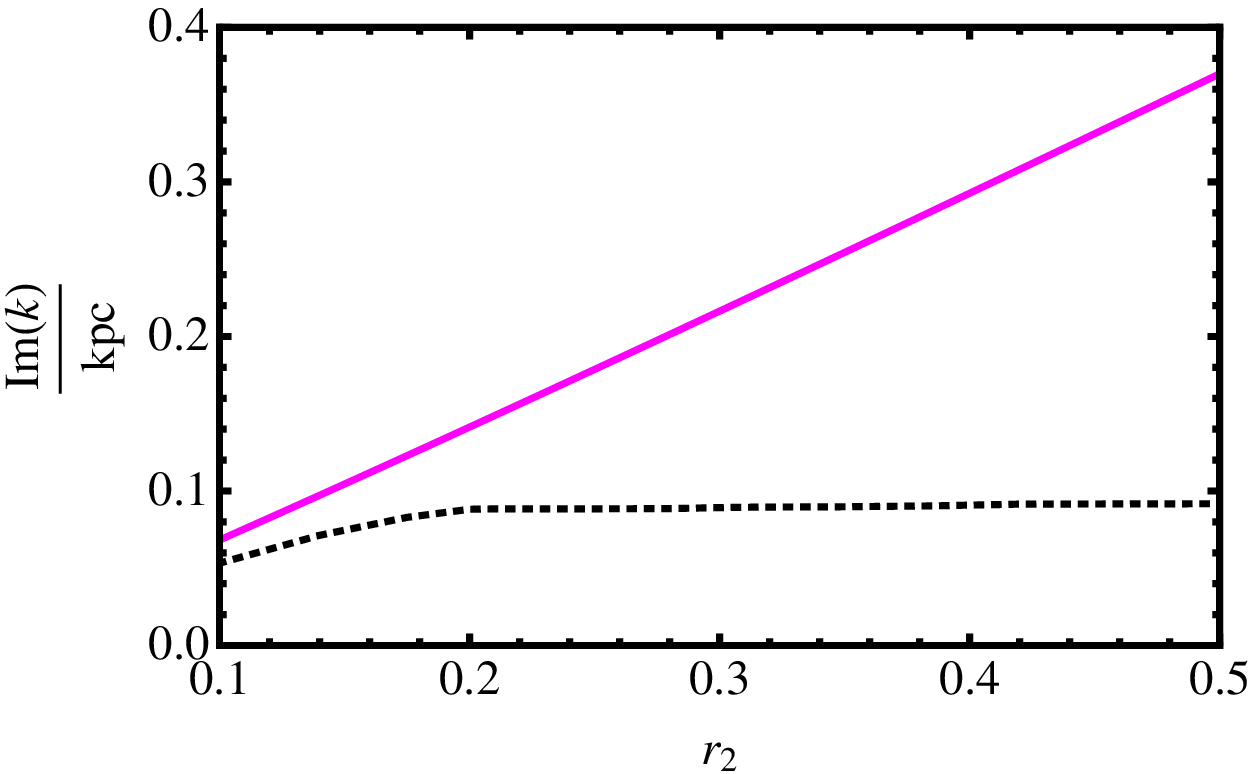}
\includegraphics[height=45mm]{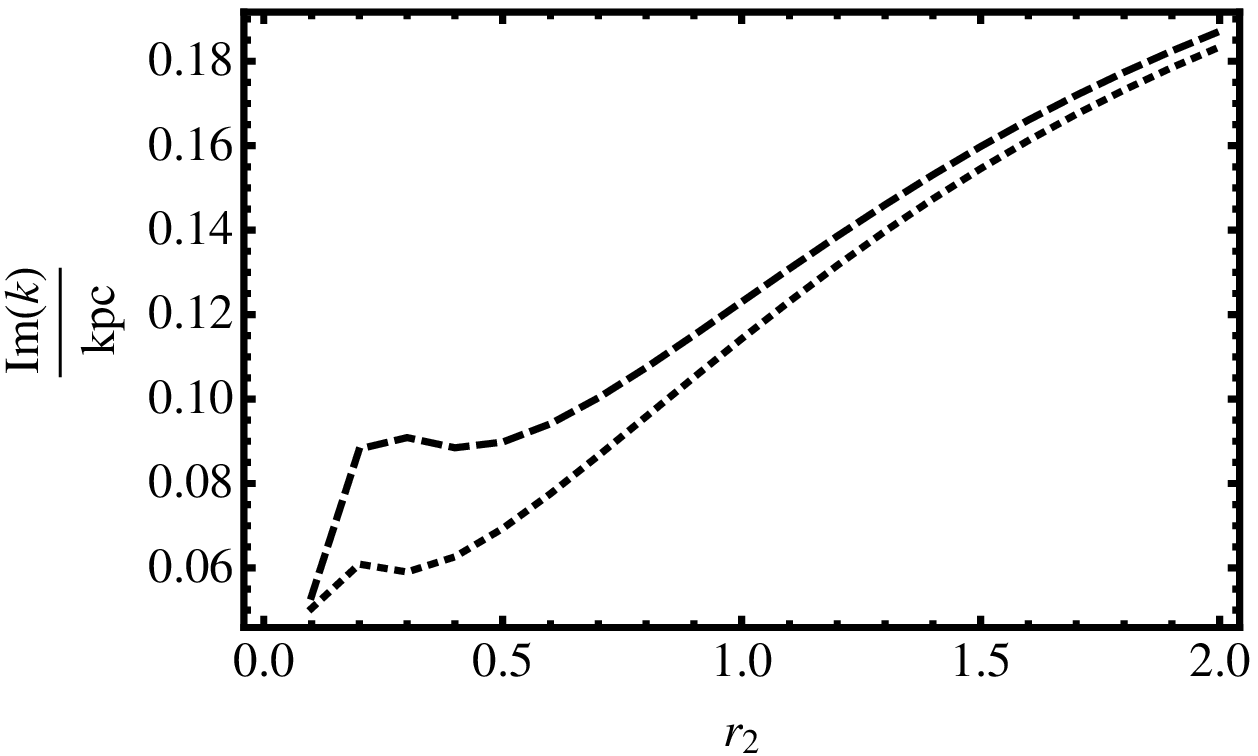}
\caption{Top: The spatial damping rates in units of kpc$^{-1}$ of acoustic waves with period 10$^7$ yr under A2199 conditions computed  according to the weak damping formula derived from eqn. (\ref{K1K0}; magenta curve) and
from the full dispersion relation (eqn. \ref{fulldr}; black dashed  curve) when all the transport coefficients have their full Coulomb values. The weak damping formula overestimates the damping rate because it assumes nearly adiabatic waves. Bottom: Comparison of the spatial damping rates with (long dashed curve) and without (short dashed curve) electron-ion thermal coupling. When coupling is turned off, electron thermal conduction reduces the electron temperature perturbation, weakening electron thermal conduction damping. }
\label{figure2aab}
\end{center}
\end{figure}

Because eqn. (1) of F05 is a weak damping formula, derived assuming the electrons are nearly adiabatic, it overestimates damping of waves in which conduction is so efficient that the electrons become isothermal. The self limiting nature of conductive damping is apparent in Fig. \ref{figure2aab}, the top panel of which compares the spatial damping rates computed derived from the weak damping formula
(eqn. \ref{K1K0}) with those derived from the full dispersion relation (eqn. \ref{fulldr}) for a wave with period 10 Myr propagating in A2199. The bottom panel compares the acoustic mode damping rate when electron-ion collisional coupling is omitted to the value when it is included. Because ion conduction is unimportant for these relatively low values of $\omega\tau_i$
(from eqn. (\ref{a2199taulambda}), $\log{\omega\tau_i}= 1.2\log{r_2}-0.259-\log{P_7}$, where
$P_7$ is the wave period in units of 10  Myr), the ions are nearly adiabatic, and electron collisions with ions prevent 
the electrons from relaxing to an isothermal state in which there is little dissipation due to electron heat conduction.

The other two modes of the system (\ref{scaledmomentum}) - (\ref{scaledienergy}) correspond to relaxation of thermal perturbations. They are nearly isobaric: $2\tiln\sim - \tildeT$. Their properties can be derived
approximately from eqn. (\ref{tildeT2})  by invoking the isobaric condition to replace $4\tiln/3$ by $-2\tilde T/3$, which leads to a quadratic equation for $K^2$. Here we give the approximate roots in the limit $\Omega\ll 1$. 

One mode, denoted by superscript $(tem)$,
 is a temperature  wave driven by electron heat conduction. It can be derived without using separate electron and ion energy equations, and has $T_{e1}\sim T_{i1}$. The dispersion relation
for the temperature mode is
\begin{equation}\label{conductionwave}
K^{(tem)2}\sim \frac{5i}{3\Omega\left(0.75\xice+1.30\xici\eps^{1/2}\right)},
\end{equation} 
or
\begin{equation}\label{conductionwave2}
\left(k^{(tem)}\li\right)^2=\frac{0.0389i\omega\tau_i}{\left(0.75\xice +0.030\xici\right)}.
\end{equation}
The temperature  mode is excited by  entropy perturbations, and damps within less than one wavelength of its source. Its characteristic wavelength is the scale on which
the heat conduction rate is comparable to the driving frequency.

The other isobaric mode, denoted by superscript $(rel)$, is driven by electron - ion 
temperature equilibration. As long as electron heat conduction is much faster than ion heat conduction, electron temperature perturbations quickly relax, so that $T_{e1}/T_{i1}
\sim{\mathcal{O}}(\eps)$. The dispersion relation for the relaxation mode is
\begin{equation}\label{relaxationwave}
K^{(rel)2}\sim -\frac{1.09\xiei}{\xici\eps^{1/2}\Omega^2}.
\end{equation}
Or,
\begin{equation}\label{relaxationwave2}
\left(k^{(rel)}\li\right)^2=-0.0254\frac{\xiei}{\xici}.
\end{equation}
The characteristic lengthscale for this wave is independent of the driving frequency, and is set by the lengthscale at which the ion thermal conduction time equals the electron - ion
relaxation time. It is excited by thermal perturbations which differ between particle species, such as  viscous heating of ions in shear flows.

We see from eqns. (\ref{conductionwave2}) and (\ref{relaxationwave2}) that both the isobaric modes are adequately described by fluid theory $(k\li\ll 1)$ as long as conduction is not too strongly suppressed.
It is clear from the large imaginary parts of the isobaric mode wavenumbers that only the acoustic wave can transport energy far from the source. The temperature and relaxation waves damp locally. 

\begin{figure}[h!]
\begin{center}
\includegraphics[height=45mm]{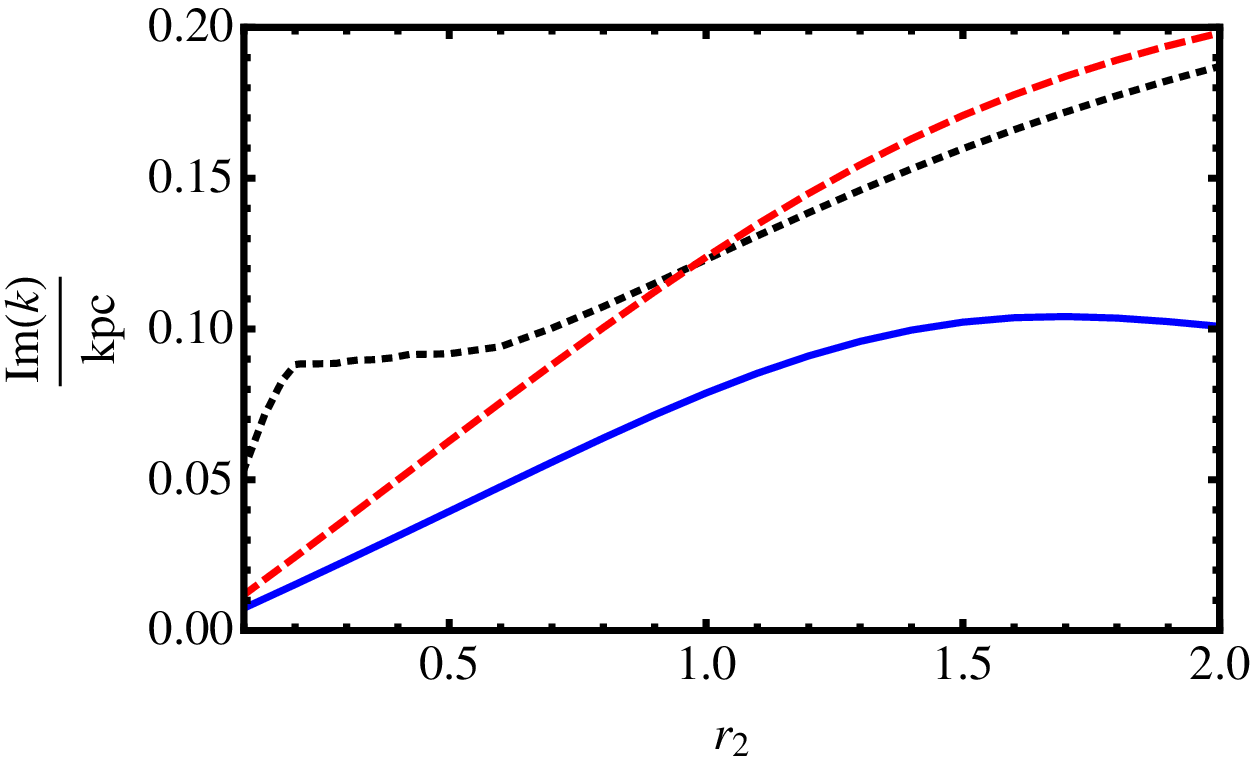}
\includegraphics[height=45mm]{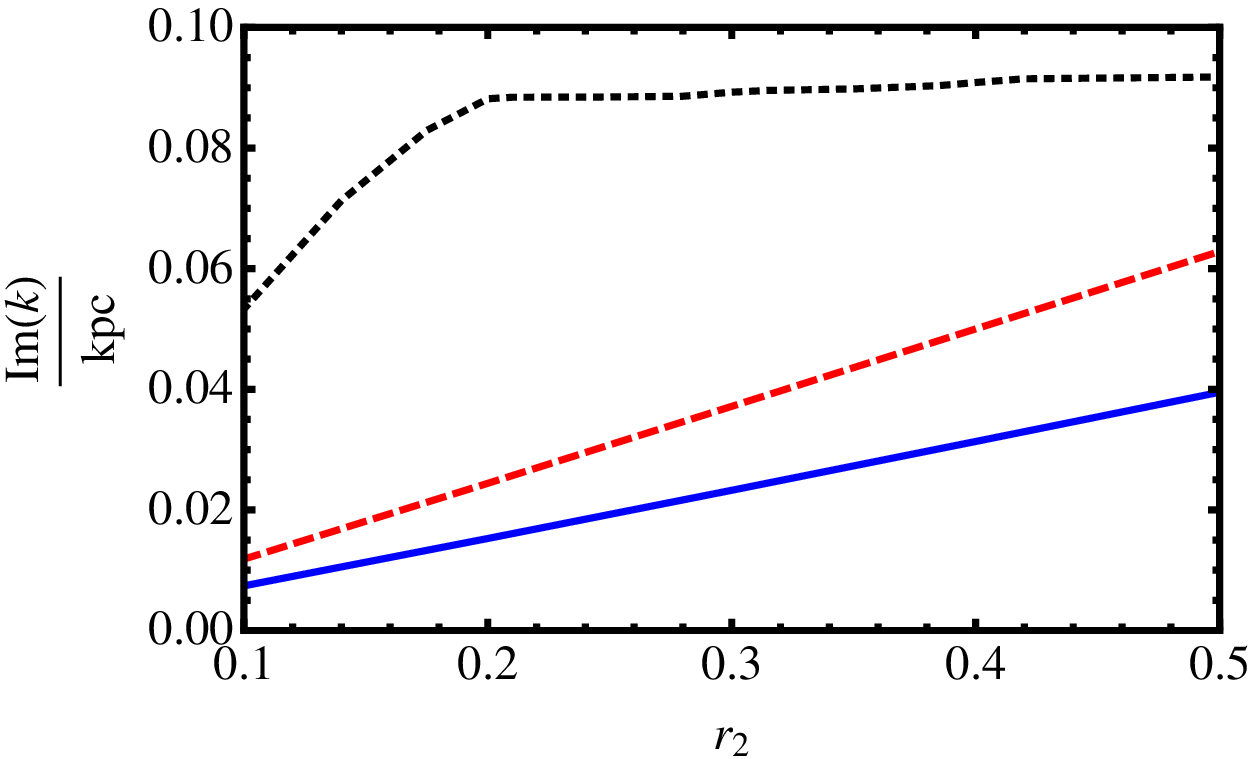}
\caption{Top: Imaginary part of the acoustic wavenumber $k$ in units of kpc$^{-1}$ for a wave with period 10 Myr as a function of position $r_2$ in A2199. The curves were computed assuming all the transport coefficients have their full Braginskii values (black dotted),  electron thermal conduction reduced to 10\% of the Braginskii value (red dashed), and both electron thermal conduction and ion viscosity reduced to 10\% of their Braginskii values (blue solid). Bottom: Same data and styles of curve for the inner 50kpc of the cluster.}
\label{figure3aab}
\end{center}
\end{figure}

Quantitative views of acoustic wave behavior for A2199 are illustrated in Figure \ref{figure3aab}. The full transport case  (black dotted curve) is the same data that was plotted in Figure (\ref{figure2aab}). The red dashed curve shows the effect of reducing $\xice$ to 0.1 while leaving other parameters the same. Although reducing the electron conductivity reduces the damping rate at small $r_2$, it delays the onset of electron isothermality, resulting in somewhat {\textit{elevated}} damping rates at
larger $r_2$. Only if both electron conductivity and ion viscosity are reduced to 0.1 their Braginskii  values is the damping rate significantly reduced (blue solid curve). 

%

Although many combinations of transport suppression parameters may be possible, there is one particular case that we wish to discuss: a model in
which ion viscosity and ion thermal conduction are completely suppressed ($\xinu=\xici\equiv 0$). This is the limit of very short ion mean free path due to
scattering by microinstabilities. It is not obvious that this model is justified in
galaxy cluster cores. According to estimates in \citet{Wieneretal2017} for the core of the Coma cluster, the ion Coulomb mean free path is short enough to suppress the firehose and mirror instabilities that drive microturbulence. However, because this model may be relevant in other environments, because it brings out the effect of electron isothermality, and because, as we show in \S\ref{ss:kinetic}, the fluid model of
ion transport overestimates the damping relative to a more accurate kinetic model,
we give results for this model here and build on them throughout the paper.
\begin{figure}[h!]
\begin{center}
\includegraphics[height=45mm]{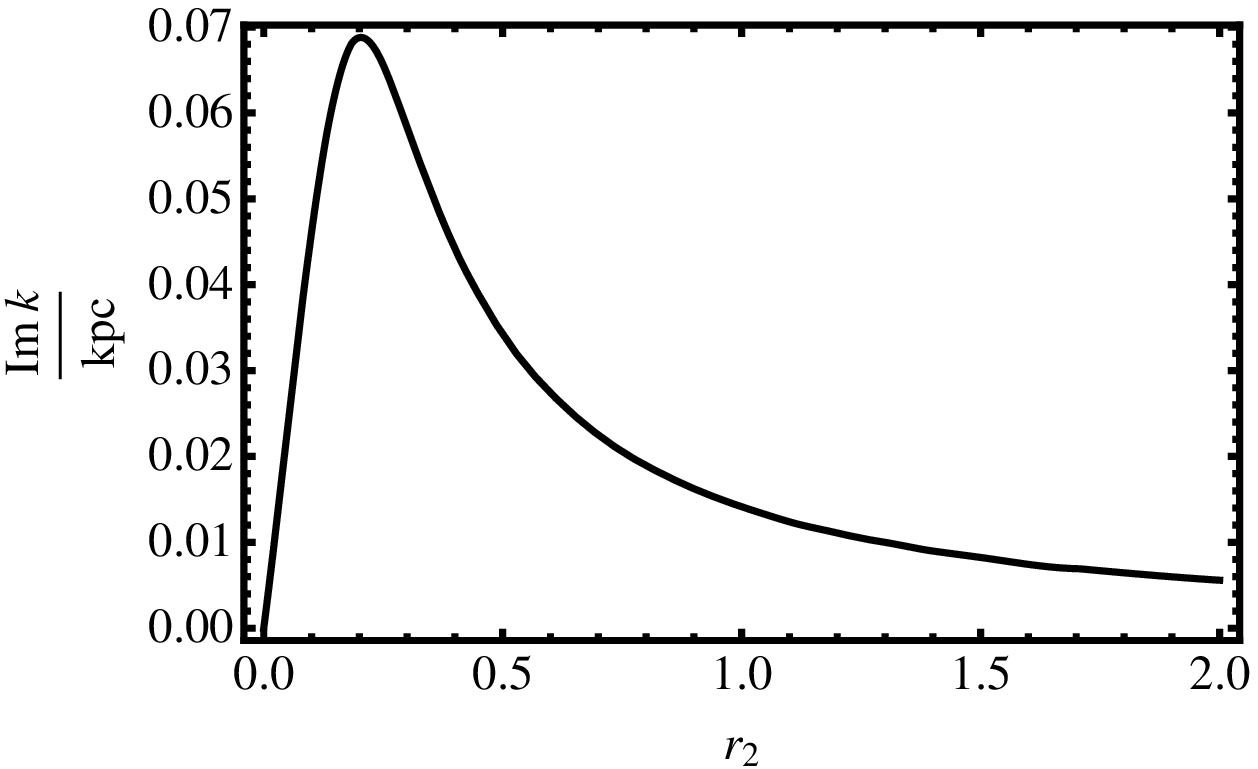}
\includegraphics[height=45mm]{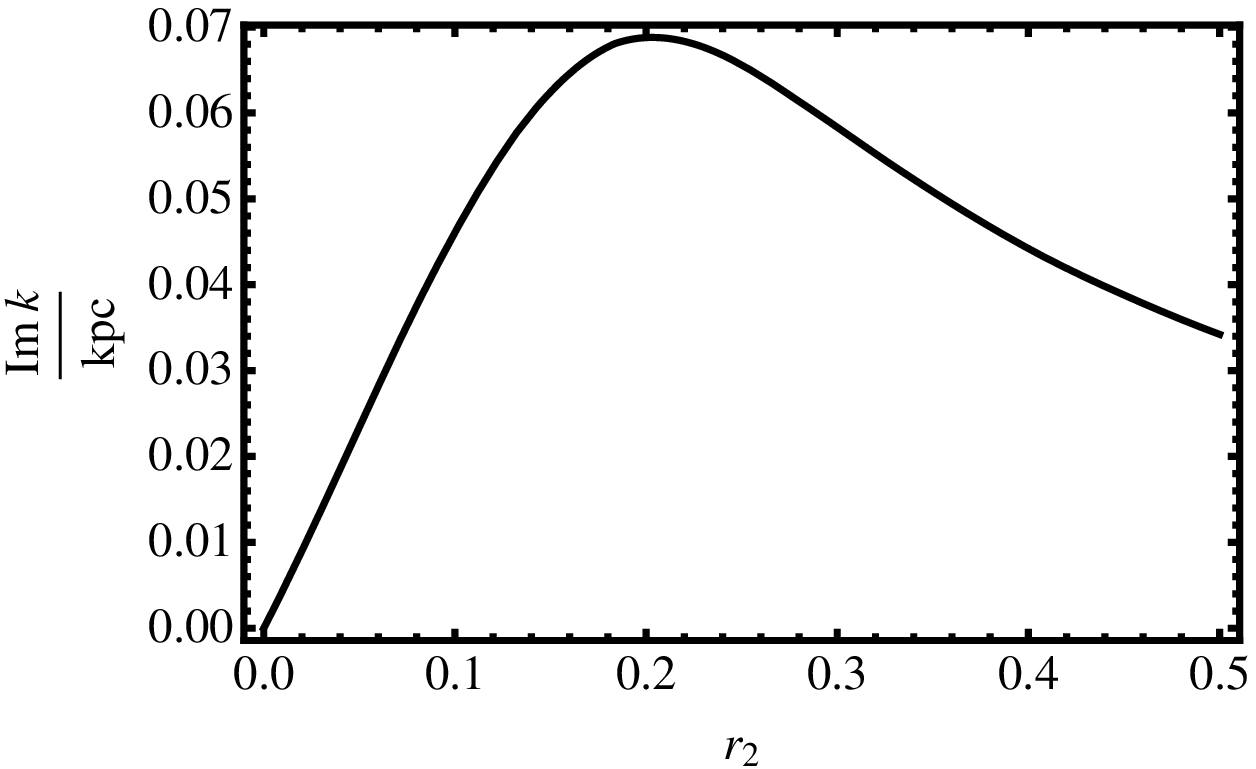}
\caption{Top: Imaginary part of the acoustic wavenumber $k$ in units of kpc$^{-1}$ for a wave with period 10 Myr as a function of position $r_2$ in A2199. The curve was computed assuming the electron thermal conductivity and electron-ion temperature equilibration rate have their full Braginskii values, but ion thermal conduction and viscosity are completely suppressed. Bottom: Same curve for the inner 50kpc of the cluster. The peak damping rate occurs at $r_2\sim 0.22$, where $\Omega\sim 3.85$. This is consistent with the estimate in
\S\ref{ss:overview} for the importance of electron thermal conduction $\omega\tau_i\sim\eps^{1/2}$ or $\Omega\sim 1$.}
\label{newnoionfig}
\end{center}
\end{figure}

The reduction in damping as the electrons become isothermal is clearly seen from
Fig \ref{newnoionfig}. The peak damping rate occurs at $r_2\sim 0.22$, where $\Omega\sim 3.85$. This is consistent with the estimate in
\S\ref{ss:overview} for the importance of electron thermal conduction $\omega\tau_i\sim\eps^{1/2}$ or $\Omega\sim 1$.
In \S\ref{s:attenuation} we will show that as a result, most of the wave attenuation and heating takes place in the inner part of the domain. 

\subsubsection{The Temperature Fluctuations}\label{sss:equilibration}

While both ions and electrons contribute to  the energy carried by waves, only $T_e$, which is generally lower than $T_i$ because of the larger electron conductivity, is observable. \cite{Fabian2006} argued for isothermal waves in Perseus, while \cite{Zhuravleva2016} found evidence for adiabatic and isobaric fluctuations as well. Here we discuss the
relationships between $\tilte$, $\tilti$, and $\tiln$ in acoustic waves. 

\begin{figure}[h!]
\begin{center}
\includegraphics[height=45mm]{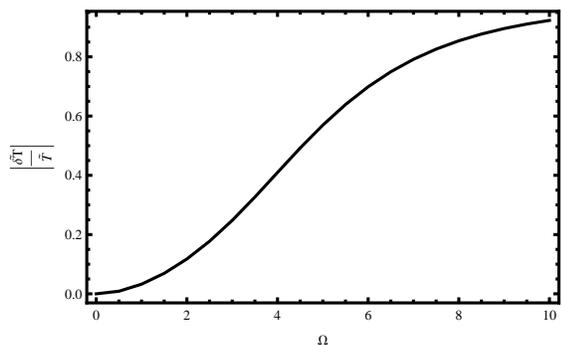}
\caption{Magnitude of  $\delta\tildeT/\tilde T$, the ratio of the difference to the sum of ion and electron temperature fluctuations defined in eqns.(\ref{tilti}) and (\ref{tilte}), as a function of scaled frequency $\Omega$ with all the
transport coefficients set to their Coulomb values. For $\Omega\ll 1$
($\omega\tau_i\ll\epsilon^{1/2}$), ions and electrons are well coupled and the temperature difference is small. As $\Omega$ increases above unity, the electrons become more isothermal and the ratio approaches unity.
For even larger $\Omega$, the ions become isothermal as well and the ratio declines again. As we will see in \S\ref{ss:kinetic}, the fluid theory is invalid for these large values of $\Omega$.}
\label{figure4a}
\end{center}
\end{figure}

The quantity $|\tilde\delta T/\tilde T|$ is plotted {\textit{vs}} $\Omega$ in Figure (\ref{figure4a}) for $0 <\Omega < 10$.  In the
 adiabatic limit ($\Omega\ll 1$), heat conduction is negligible and  both $\tilte$ and $\tilti$ are related to the $\tiln$ by the usual adiabatic relation $\tilte=\tilti=2\tiln/3$ for
 ideal gases. Electron heat conduction  becomes more important as $\Omega$ increases away from zero, with $|\tilde\delta T/\tilde T|$ being of order
 $\Omega^2$ for $\Omega\ll 1$. Differences between $\tilte$ and $\tilti$ become significant for moderate $\Omega$ as the electrons become isothermal while the ions remain nearly adiabatic; isothermal electrons
 but adiabatic ions corresponds to $|\tilde\delta T/\tilde T|\rightarrow 1$. At large $\Omega$ 
  the fluid theory should be replaced by a semicollisionless or collisionless theory for these large values of $\Omega$ (\S\ref{ss:kinetic}), so we do not extend the plot to large values here.   

\begin{figure}[h!]
\begin{center}
\includegraphics[height=45mm]{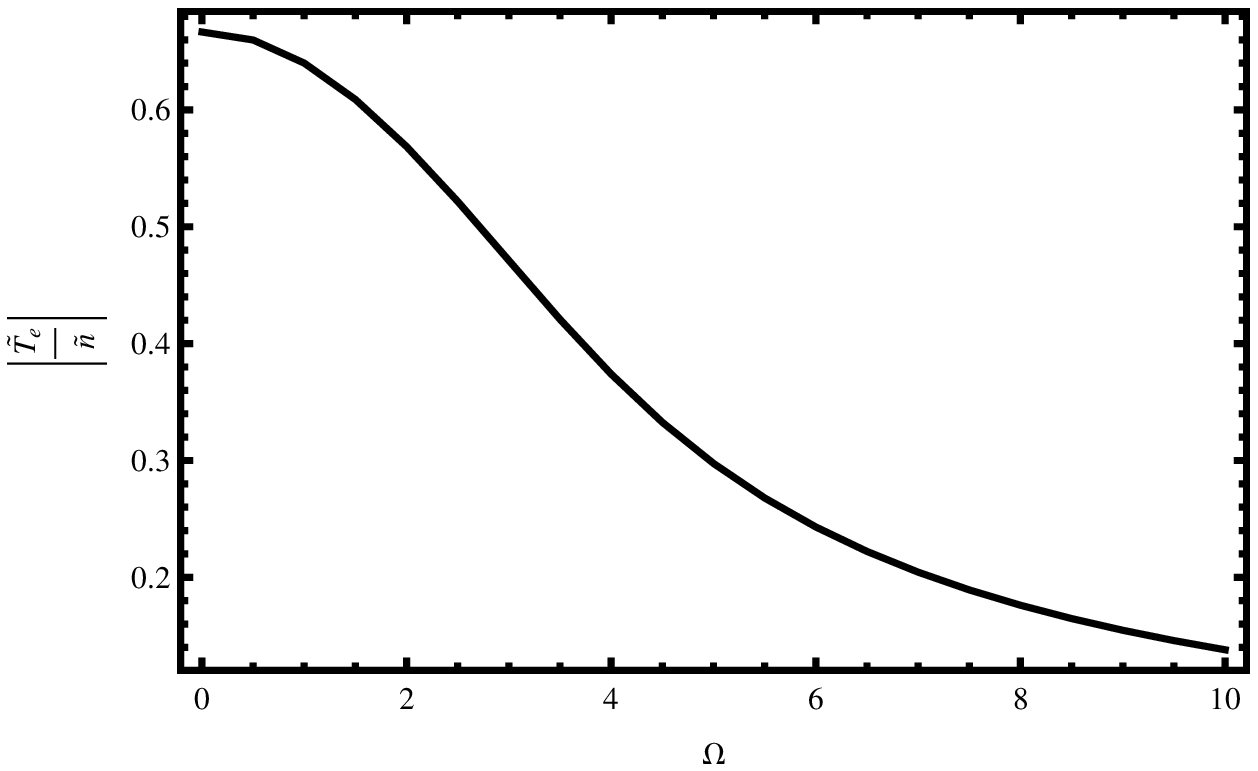}
\includegraphics[height=45mm]{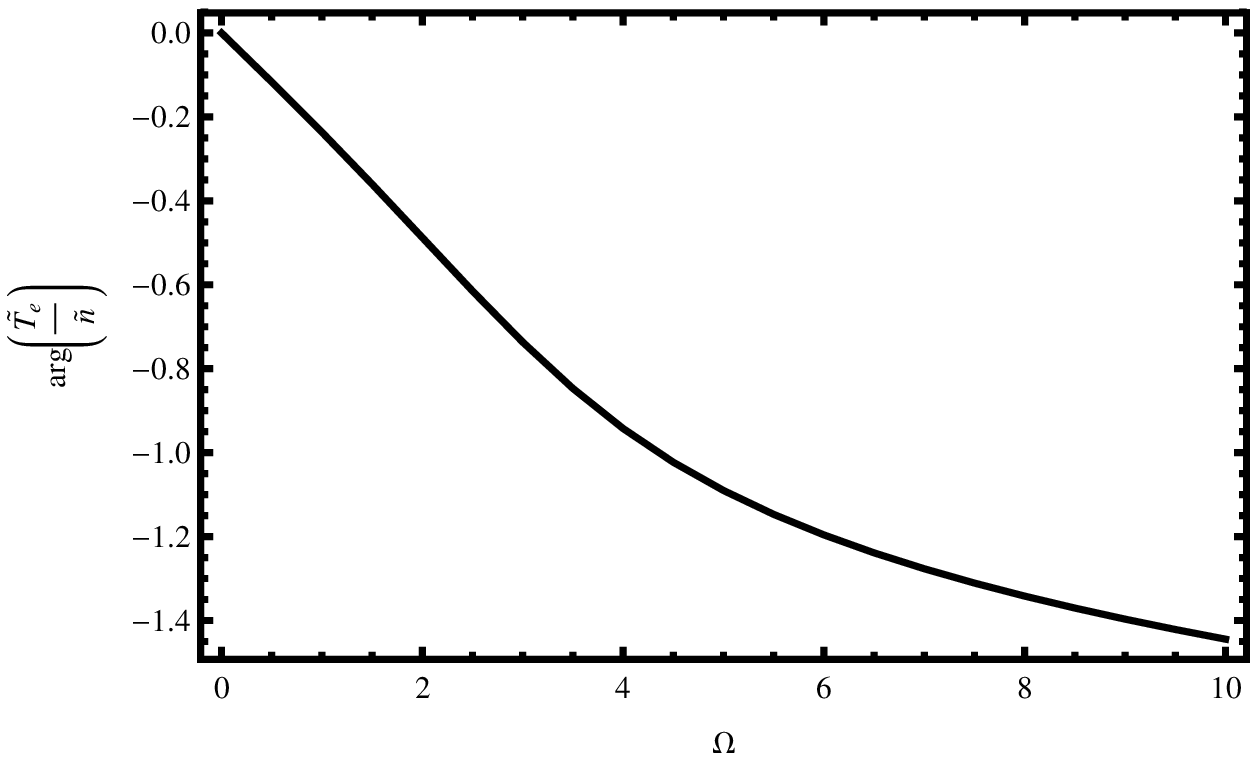}
\caption{Top: The absolute value of the ratio of the normalized electron temperature perturbation to the normalized density perturbation, $\vert\tilte/\tiln\vert$ {\textit{vs}} the scaled frequency $\Omega$.  For  $\Omega
\ll 1$ the ratio has the usual adiabatic value of $2/3$, but decreases to zero as conduction becomes important at high frequencies, where the electrons are isothermal. Bottom: The phase of  $\tilte/\tiln$ {\textit{vs}}
$\Omega$. As expected for a damped wave, $\tilte$ always lags $\tiln$; the lag increases with increasing $\Omega$, but the accompanying decrease in $\vert\tilte/\tiln\vert$ may make the phase lag difficult to detect. }
\label{figure5aab}
\end{center}
\end{figure}
%
 
The  magnitude and phase of $\tilte$ relative to 
$\tiln$ are plotted {\textit{vs}} $\Omega$ in Figure (\ref{figure5aab}).  For $\Omega\ll 1$ the waves are almost adiabatic, $\tilte/\tiln\sim 2/3$ and the quantities are almost in phase. As
$\Omega$ increases the relative amplitude of $\tilte$ decreases due to the increasing importance of conduction and $\tilte$ lags $\tiln$ in phase by an increasing amount. The lag is expected for a damped wave; it means
that the fluid is losing heat at the time of greatest compression (similar to the damped version of the classic Eddington valve invoked to explain self excited stellar pulsations). However, the phase shift may be
difficult to observe owing to the small  amplitude of $\tilte$ relative to $\tiln$ in the range of $\Omega$ where the phase shift is large.

\subsection{Kinetic Theory}\label{ss:kinetic}

%

In the collisionless limit ($\kli\rightarrow\infty $), ion acoustic waves are described by kinetic theory.  
The
electrons are isothermal ($\gamma_e = 1$) and the ions are adiabatic, with one degree of freedom ($\gamma_i = 3$). Collisional damping processes are negligible, and dissipation is 
primarily due to ion
Landau damping: the absorption of wave energy by ions traveling at slightly less than the speed of the wave (electron Landau damping is weaker by a factor of
$\eps^{1/2}$). Because the wave speed is near the ion thermal speed (unless $T_e\gg T_i$), collisionless damping is strong. The dispersion relation in dimensionless form written for
real $k$ and complex $\omega$ is
\begin{equation}\label{kineticdr}
\omega\tau_i=\left(2-0.85i\right)k\li
\end{equation} 
for $T_e=T_i$. 
According to eqn. (\ref{kineticdr}), ion Landau damping reduces the wave amplitude to 7\% of its initial value within one wavelength: ion acoustic waves
essentially cannot
propagate in a collisionless plasma with $T_e=T_i$. 
\begin{figure}[h!]
\begin{center}
\includegraphics[height=45mm]{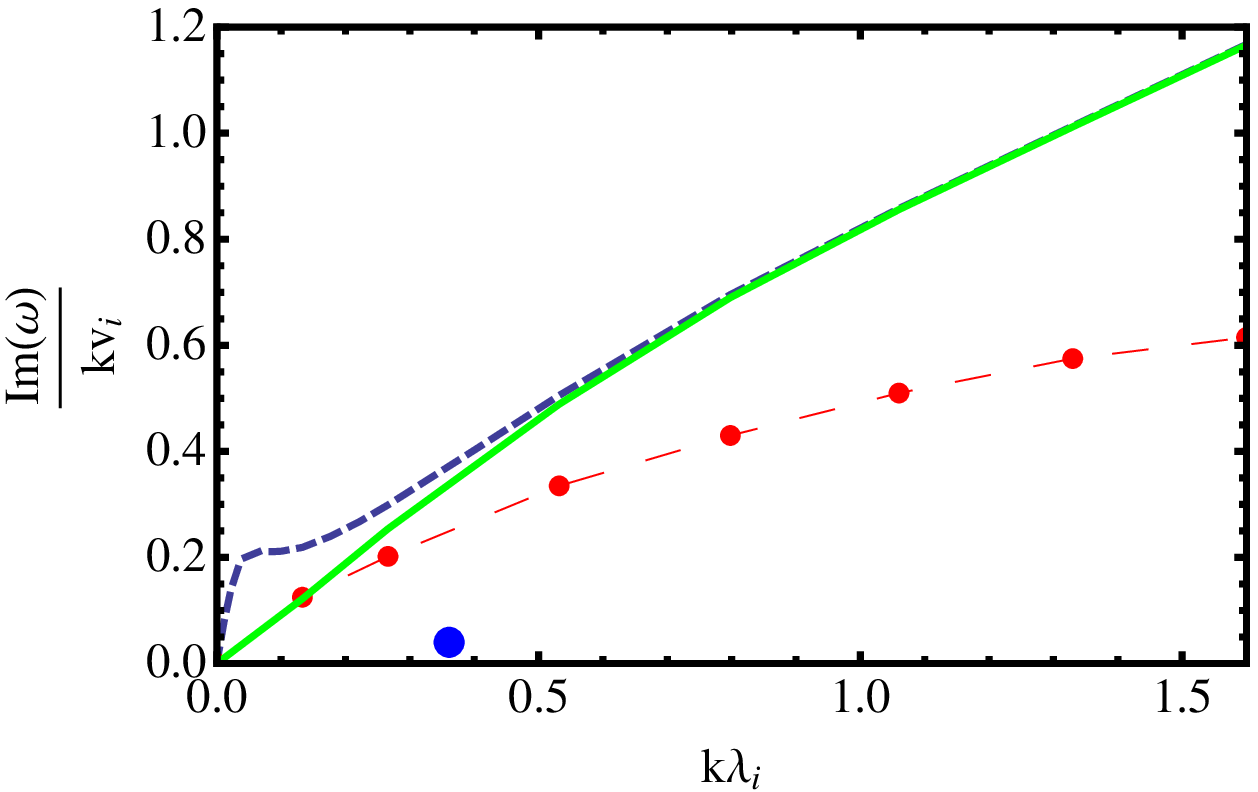}
\includegraphics[height=45mm]{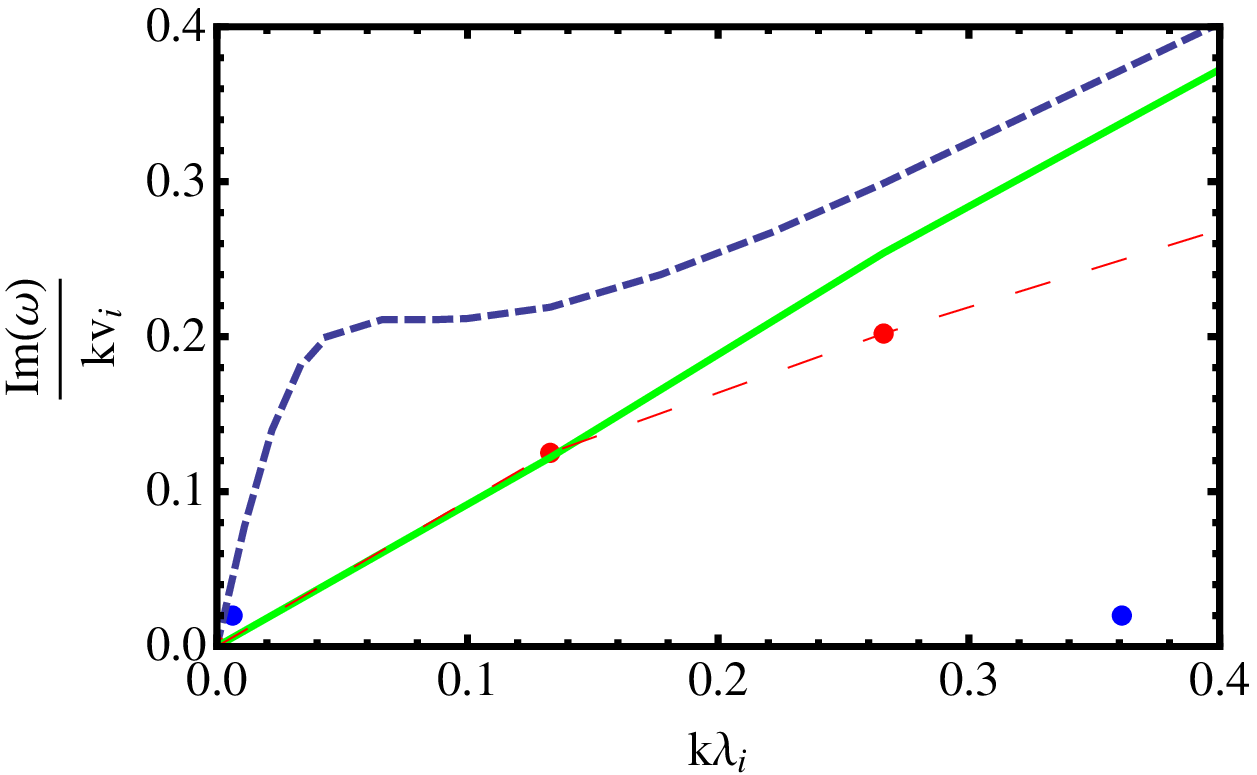}
\caption{Comparison of the scaled wave damping rate $Im[\omega]/kv_i$ as a function of $k\lambda_i$ in three different treatments. The top and bottom  panels show the same
data over different ranges. The black dashed curve is the full fluid dispersion relation. The green curve was obtained from the fluid theory  assuming isothermal electrons ($\tilte\equiv 0$) and ignoring electron-ion collisional coupling because of its relative slowness. The red points are the damping rates taken from Table 2 of \citet{KulsrudOno1975} and are based on solving the Fokker-Planck equation for ions accounting for ion-ion collisions only and assuming isothermal electrons. We have added the long dashed curve  to facilitate comparison with the other curves. The blue points show $k\lambda_i$ for A2199 at 5 and 100 kpc from the cluster center. The elevated damping rates at low $\kli$ seen in the full fluid theory are due
to electron thermal conduction. As $\kli$ and $\kle$, which is almost the same, increase, the electrons become more isothermal and the damping rates predicted by the green and black dashed curves converge. On the other hand, beyond $\kli\sim 0.2$, the fluid theory significantly overestimates the damping rate.} 
\label{figure6aab}
\end{center}
\end{figure}
%

 
\cite{KulsrudOno1975} studied the transition from collisional to collisionless behavior by solving the linearized Boltzmann equation with a Fokker-Planck collision
operator. They considered only ion-ion collisions, and assumed the electrons are isothermal. Therefore, the damping rates they calculated account for ion viscosity and ion thermal 
conduction, but not electron thermal conduction or electron-ion temperature relaxation. 

In Figure \ref{figure6aab}, the results of the Fokker-Planck calculation from Table 2 of \cite{KulsrudOno1975} (red points connected by a dashed curve to facilitate comparison with fluid models) are compared
to the full fluid model (black dashed curve) and a fluid model with isothermal electrons and electron-ion coupling switched off (solid green curve). Note that we have followed \citet{KulsrudOno1975} and computed temporal, not spatial, damping rates. The convergence of the two fluid curves shows that the isothermal electron-adiabatic ion model captures the fluid behavior for  $k\li <\sim 0.2$. The large blue points mark the values of $k\li$ at 5 kpc and 100 kpc from the center of A2199. Within this range, the Fokker-Planck and fluid formulae
 agree to within 20\%. Considering the differences between the physical models this is reasonably good agreement and shows that collisionless effects are
small. Comparison of Figure \ref{figure1aab} with Figure 2 of \citet{KulsrudOno1975} suggests that the fluid description is adequate within the inner 100 - 150 kpc of galaxy clusters where acoustic waves
are observed.

While we have focused here on the inner parts of the ICM, density and temperature profiles have been measured at larger radii as well. 
Although both $n_e$ and $T$ generally decline  with $r$, the entropy parameter $T/n_e^{2/3}$
is found to increase with $r$ in a large sample of clusters
\citep{Pratt2010}, and $\lambda_i$ (which is proportional to $T^2/n$) appears to do so as well. For example, the $n_e$ and $T$ profiles derived by Simionescu et al. (2012) for Perseus
give $\lambda_i\sim$ 20 - 25 kpc at $r_2\sim$ 10. This suggests that acoustic waves launched by dynamical disturbances will damp almost immediately in the outer parts of massive galaxy clusters. Lower mass clusters are cooler and more collisional; e.g. the profiles derived for Centaurus by \cite{Walker2013} give $\lambda_i
\sim$ 4 kpc at $r_2\sim$ 10, indicating a more favorable environment for the propagation of  waves.  

 Collisionless acoustic wave damping heats the ions, not the electrons. Using eqn. (\ref{tei}) and the relation $\tau_{ei}=\tau_e/\eps$ we see that if
$T\sim$ 5 keV and $n_e\sim$ 10$^{-4}$ cm$^{-3}$, $\tau_{ei}\sim$ 1.6 Gyr. This suggests that $T_i$ may exceed $T_e$ in the outer parts of galaxy clusters, making
 collisionless damping
even stronger. It further suggests that pressure models based in $T_e$ may underestimate the ion pressure.

%% file: waves_10302017_Sections45.tex
\section{Wave Attenuation}\label{s:attenuation}

Damping attenuates wave amplitude by a factor $\exp{(-{\mathcal{A}}(a,r))}$
\begin{equation}\label{A1}
{\mathcal{A}}(a,r)\equiv\int_{a}^{r}k_idr^{'},
\end{equation}
for a wave launched at $a$. Equation (\ref{A1}) can be written in terms of the scaled variables as
\begin{equation}\label{A2}
{\mathcal{A}}(\Omega_a,\Omega_r)=\epsilon^{1/2}\int_{\Omega_a}^{\Omega_r}\frac{K_id\Omega}{v_id\tau_i/dr},
\end{equation}
where $\Omega_{a,r}$ are the values of $\Omega$ at $a$ and $r$. If the temperature and density are power laws, $(T,n)\propto (r^{\alpha},r^{-\beta})$, we can write $(v_i,\tau_i)=(v_{i0}r_{2}^{\alpha/2},\tau_{i0}r_{2}^{(3\alpha +2\beta)/2})$. Equation (\ref{A2}) is then
\begin{equation}\label{A3}
{\mathcal{A}}(\Omega_a,\Omega_r)=\frac{2}{3\alpha+2\beta}\frac{r}{r_2}\frac{\epsilon^{1/2}}{v_{i0}\tau_{i0}}\left(\frac{\omega\tau_{i0}}{\epsilon^{1/2}}\right)^q\int_{\Omega_a}^{\Omega_r}\frac{K_id\Omega}{\Omega^q},
\end{equation}
where $q\equiv 2(2\alpha+\beta -1)/(3\alpha + 2\beta)$ and we have used
\begin{equation}\label{r2O}
r_2=\left(\frac{\epsilon^{1/2}\Omega}{\omega\tau_{i0}}\right)^{2/(3\alpha + 2\beta)}.
\end{equation}

For A2199,
profiles of {\cite{Johnstone2002}}: $\alpha = 0.3$,
$\beta = 0.75$, $q = 0.29$.
\begin{figure}
\begin{center}
\includegraphics[height=45mm]{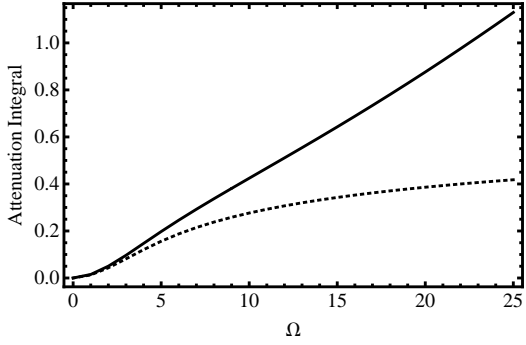}
\caption{Attenuation integrals $\int_{0.001}^{\Omega} K_i(S)S^{-q}dS$ for $q=0.29$, the value derived for A2199,
for two transport models. The solid curve represents full Braginskii. The dotted curve shows the effect of completely
suppressing ion transport, bringing out the effect of the
transition to electron isothermality. 
}
\label{new_attenuation1}
\end{center}
\end{figure}
The integral in eqn. (\ref{A3}) is plotted in Fig. \ref{new_attenuation1} for $\Omega(a) = 0.001$ (for all practical purposes, the center) out to $\Omega =
25$. 

\begin{figure}
\begin{center}
\includegraphics[height=45mm]{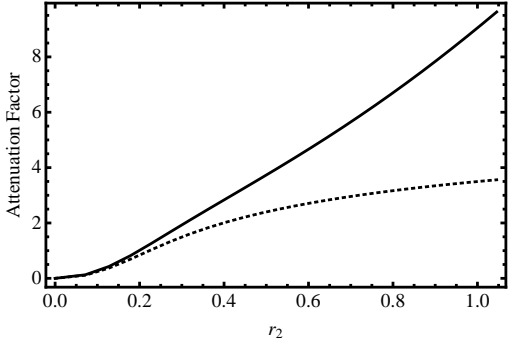}
\caption{Attenuation factors ${\mathcal{A}}$ defined in eqns.
(\ref{A1}) - (\ref{A3})
for two transport models. The solid curve represents full Braginskii. The dotted curve shows the effect of completely
suppressing ion transport, bringing out the effect of the
transition to electron isothermality (collisional electron - ion temperature coupling is present in the model).}
\label{new_attenuation2}
\end{center}
\end{figure}

With
$\tau_{i0}=2.76\times 10^{13}$s,
$v_{i0}=6.43\times 10^7$cm s$^{-1}$, the prefactor multiplying
the integral in eqn. (\ref{A3}) is $8.52/P_7^{0.29}$, where
$P_7$ is the wave period in units of 10 Myr, and
\begin{equation}\label{O2199}
\Omega=\frac{23.7}{P_7}r_2^{1.2}.
\end{equation}
%
The attenuation factor ${\mathcal{A}}$ for a wave with $P_7=1$ is plotted {\textit{vs}} $r_2$ in Fig. \ref{new_attenuation2},

 The sensitivity of the attenuation factor to gas temperature depends on the transport model and cluster profiles. If we denote the temperature at the reference level where $\tau_{i0}$ is measured by $T_0$, then the prefactor in eqn. (\ref{A3}) scales as $T_{0}^{(3q-4)/2}$, or
-1.57 for A2199. For a 10\% increase in $T_0$, this reduces the prefactor to 7.34 for waves with $P_7=1$.
On the other hand, $\Omega$ scales as $T_0^{3/2}$. For A2199, increasing $T_0$ by 10\%
increases $\Omega(0.75)$ for a wave with $P_7=1$ to 19.4, increasing  ${\mathcal{A}}$ from 6.18 to 6.22 in the full Braginskii case and decreasing ${\mathcal{A}}$ from 3.07 to 2.80 in the full suppression of ion transport case.

If {\textit{all}} forms of transport are so strongly suppressed that the weak damping formula (eqn. \ref{K1K0}) applies the attenuation factor can be written
in the form
\begin{equation}\label{weakattenuation}
{\mathcal{A}}(0,r_2)=607\left(\frac{10 Myr}{P}\right)^2\xi_{tot}r_2^{2.05},
\end{equation}
where $\xi_{tot}\equiv 0.0048\xinu +0.045\xice + 0.0018\xici$ is a total suppression factor. If all transport coefficients have their full Braginskii values, then $\xi_{tot}=0.0516$. Equation (\ref{weakattenuation}) can be used to solve for the degree of transport suppression needed to achieve any given attentuation factor. For example, if $P$ = 10 Myr then ${\mathcal{A}}(0,0.5)=1$ if $\xi_{tot}=0.0068$, or 13\% of the Braginskii value. 
\section{Heating by Wave Dissipation}\label{s:heating}


The heating rate is given by F05 in terms of an acoustic luminosity $L_s(r)$ which takes the value $L_{inj}$ at an injection radius $r_{inj}$
 is attenuated by dissipation as it travels through the medium according to 
\begin{equation}\label{Lseqn}
\frac{dL_s}{dr}=-2k_iL_s,
\end{equation}
 The heating rate per unit volume is 
\begin{equation}\label{ediss}
\epsilon_{diss}= 2k_i\frac{L_s(r)}{4\pi r^2}.
\end{equation} 
If we define a wave energy flux $F_s$ such that $L_s=4\pi r^2 F_s$, then eqns. (\ref{Lseqn}) and (\ref{ediss}) identify the heating rate with the divergence of $F_s$. Since $k_i\equiv 0$ for an ideal system, the heating rate vanishes without dissipation.

Equation (\ref{ediss}) is intuitively plausible, but it has two limitations. One is that it does not separate energy input due to PdV work from energy input due to heating. The other is that it can be estimated from, e.g., our understanding of AGN power output (as was done in F05), but not measured directly within the ICM.

To illustrate the first problem, we start with the  energy conservation law for the combined electron - ion fluids (B65)
%
\begin{equation}\label{energyconservation}
\frac{\partial\epsilon}{\partial t}=-\mbfnabla\cdot\mbfF,
\end{equation}
where
\begin{equation}\label{energydensity} 
\epsilon\equiv\frac{1}{2}\rho u^2 +\frac{3}{2}\left(P_e+P_i\right) 
\end{equation}
 is the combined mechanical and thermal electron and ion energy density and 
\begin{equation}\label{energyflux}
\mbfF\equiv\left(\epsilon + P_e+P_i\right)\mbfu+\xinu\mbfpi\cdot\mbfu-\xice\chi_e\mbfnabla T_e -\xici\chi_i\mbfnabla T_i
\end{equation}
 is a generalized energy flux made up of the mechanical, enthalpy, viscous, and conductive 
energy fluxes, summed over electrons and ions. 

Equation (\ref{energyflux}) can be used to calculate $L_{inj}$, the rate at which an oscillating source transmits energy to the surrounding medium. For simplicity we consider a spherically symmetric source oscillating around an equilibrium position at $r=a$.  By symmetry, the energy flux is radial: 
 $\mbfF = \hat r F$. The energy per solid angle outside the source changes according to 
\begin{multline}\label{transmit}
\frac{\partial}{\partial t}\int_{a}^{\infty}\epsilon r^2dr=\int_{a}^{\infty}\frac{\partial\epsilon}{\partial t}r^2dr-\epsilon(a)a^2\frac{\partial a}{\partial t}\\ =F(a)a^2 - \epsilon(a)a^2u(a),
\end{multline} 
where in the second equality we have used eqn. (\ref{energyconservation}), assumed $F$ decays faster than $r^{-2}$ at infinity, as must occur for a damped wave, and identified $\partial a/\partial t$ with $u(a)$. Using eqn. (\ref{energyflux}) we then find for 
the rate of energy input to
the medium
\begin{multline}\label{transmit2}
\frac{\partial}{\partial t}\int_{a}^{\infty}\epsilon r^2 dr \\ =a^2\left[
(P_e+P_i)u+\xinu(\mbfpi\cdot u)_r-\xice\chi_e\frac{\partial T_e}{\partial r}-\xici\chi_i\frac{\partial T_i}{\partial r}\right]_a.
\end{multline}
Equation (\ref{transmit2}) shows that the global energy is changed by $PdV$ work (the first term on the right hand side of eqn. \ref{transmit2}) and by dissipation (the viscosity and heat conduction terms). The wave contribution to eqn. (\ref{transmit2}) comes from expanding $F$ to
second order in the wave amplitude. We will not do that here, except to note that in an ideal medium the pressure and velocity perturbations
are out of phase by $\pi/2$, so only a damped or growing wave can do work on its environment
\citep{Goldreich1989}.

In order to eliminate the ambiguity between doing work and adding heat, we appeal to the entropy conservation law (B65, \cite{LandauLifshitz1987})

\begin{equation}\label{entropyeqn}
\frac{\partial S}{\partial t}+\mbfnabla\cdot\left(S\mbfV +\frac{\mbfq_e +\mbfq_i}{T}\right)=\theta ,
\end{equation}
where $S$ is the plasma entropy per volume, $\mbfq_e$, $\mbfq_i$ are the electron and ion heat fluxes, and $\theta T\equiv T(\theta_{\nu}  + \theta_{ce} +
\theta_{ci} +\theta_{ei})$ are the rates of entropy production per volume due to ion viscosity, electron heat conduction, ion heat conduction, and electron - ion collisional
heat exchange, respectively. For waves, these heat sources take the forms
\begin{equation}\label{tthetav}
T\theta_{\nu}=-\frac{1}{2}\xinu\pi_{ab}W_{ab} = 1.28\xinu\frac{\rho v_i^2}{\tau_i}\left(\omega\tau_i\right)^2\vert\tiln\vert^2,
\end{equation}
\begin{equation}\label{tthetaec}
T\theta_{ce}=\frac{n\xice\chi_e}{T}\vert\nabla T_{e1}\vert^2=\frac{2.23}{\eps^{1/2}}\xice\frac{\rho v_i^2}{\tau_i}\vert k\li\vert^2\vert\tilte\vert^2,
\end{equation}
\begin{equation}\label{tthetaic}
T\theta_{ci}=\frac{n\xici\chi_i}{T}\vert\nabla T_{i1}\vert^2 = 3.90\xici \frac{\rho v_i^2}{\tau_i}\vert k\li\vert^2\vert\tilti\vert^2,
\end{equation}
\begin{equation}\label{tthetaei}
T\theta_{ei}=3\xi_{ei}\frac{m}{M}\frac{n}{T\tau_e}\left( T_{e1} - T_{i1}\right)^2 = 4.24\xi_{ei}\eps^{1/2}\frac{\rho v_i^2}{\tau_i}\vert\tilte - \tilti\vert^2
\end{equation}
where the notation comes from \S 2.  

\begin{figure}
\begin{center}
\includegraphics[height=45mm]{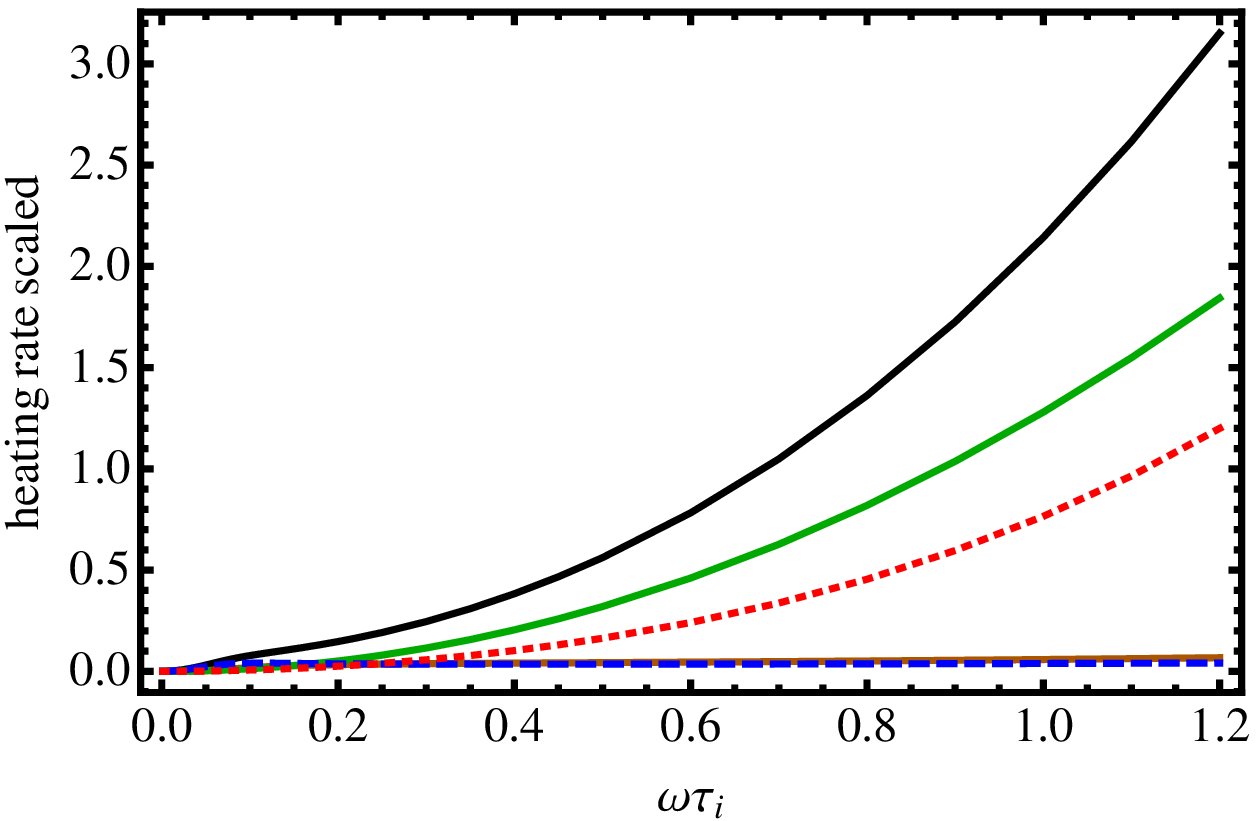}
\includegraphics[height=45mm]{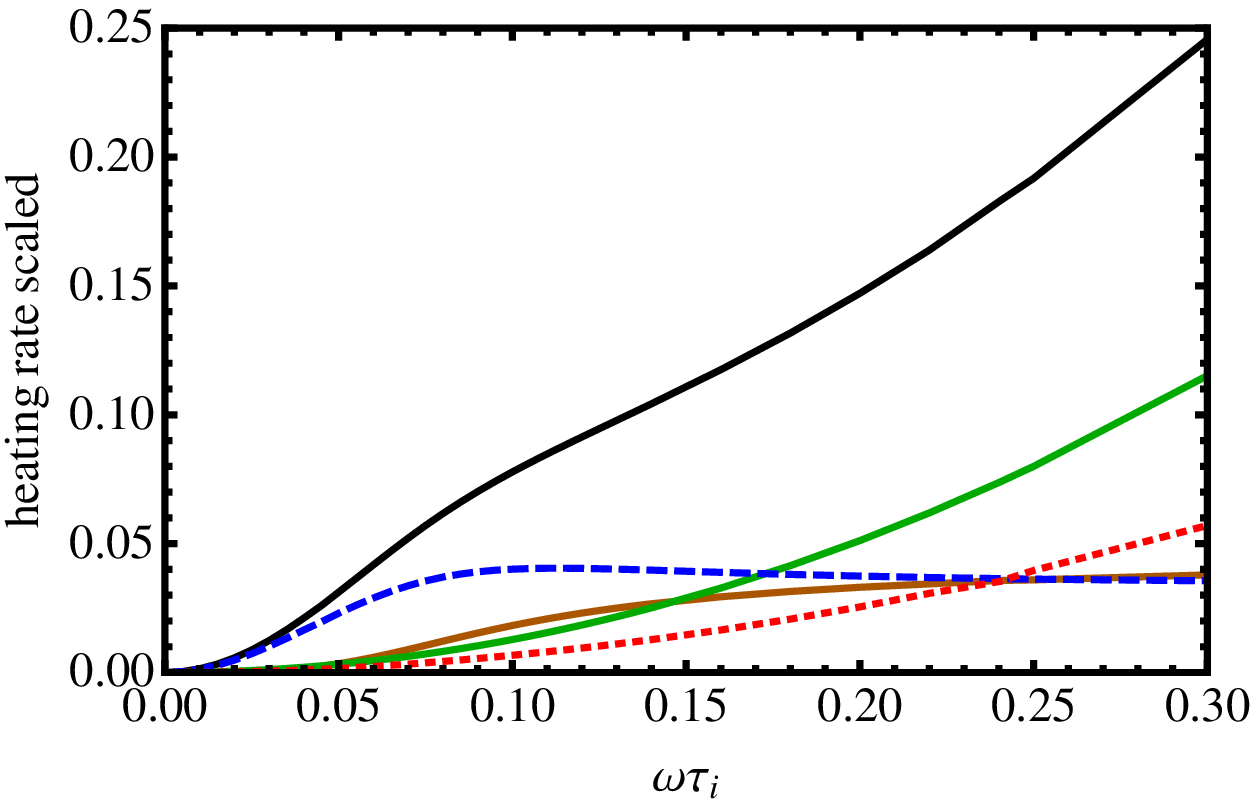}
\includegraphics[height=45mm]{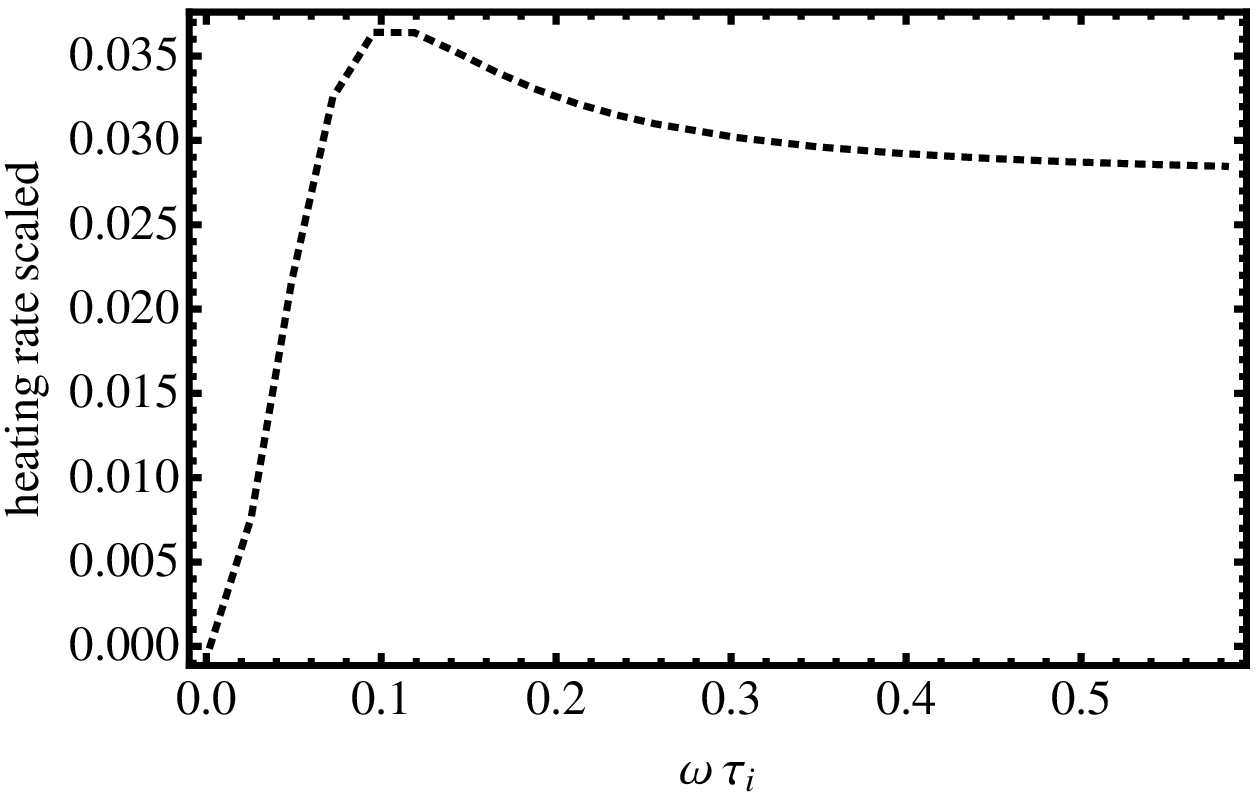}
\caption{Top and middle panels: Scaled dimensionless heating rates due to ion viscosity (solid green curve), ion heat conduction (dotted red curve), electron heat conduction (blue dashed curve), and electron-ion thermal coupling (solid orange curve) from eqns. (\ref{tthetav})-(\ref{tthetaei}) and their sum, the function $\Psi$, (solid black curve) as functions of the collisionality parameter  $\omega\tau_i$ when all transport coefficients have their full values. To convert these rates to energy per volume per time for a wave with density amplitude $\tiln$, multiply by $\rho v_i^2/\tau_i\tiln^2$. The top and middle panels plot the same data, but the middle panel is zoomed in. The points $\omega\tau_i = 1.2$ and $\omega\tau_i=0.3$ correspond to $r_2=1.92$ and $r_2 = 0.60$ respectively for a wave with $P=10$Myr. The bottom panel shows the
scaled heating rate due to electron thermal conduction when ion transport is completely suppressed. It is very similar to the electron heating rate in the full Braginskii transport case.}
\label{figure7aab}
\end{center}
\end{figure}

The contributions of viscosity, electron and ion heat conduction, and electron-ion thermal coupling to the heating rate computed from eqns. (\ref{tthetav}) - (\ref{tthetaei}) are shown
as functions of $\omega\tau_{i}$ in top and middle panels of Figure \ref{figure7aab} for the case that all transport coefficients have their full Coulomb values. The total scaled heating rate summed over all contributions is given by the black curve.
At small $\omega\tau_i$ electron thermal
conduction dominates, but is overtaken by ion heat conduction and viscosity as $\omega\tau_i$ increases beyond a few tenths, mirroring the contributions of these processes to damping. As the electrons approach isothermality, their temperature perturbation $\tilte$ is determined by balancing thermal conduction against compression. This leads to $\tilte\propto\tiln/k$ and $\nabla\tilte$ independent of $k$, giving a nearly constant rate of entropy production.  Ion thermal conduction is relatively unimportant for the range of $\omega\tau_i$ considered here, so entropy production is nearly quadratic in $\omega\tau_i$ (or $k$), as is viscous heating. 
The bottom panel of Fig. \ref{figure7aab} shows the scaled entropy production rate by electron thermal conduction when ion transport is completely suppressed. It is very similar to the blue dashed curves in the
top and middle panels. The location of the peak coincides with the maximum damping rate, which
is near $r_2\sim 0.22$ (see Fig. \ref{newnoionfig}).

Although these results are presented in dimensionless form, they are readily converted to physical values. Expressing $\tilte$ and $\tilti$ in terms of $\tiln$ using eqns. (\ref{tilti}), (\ref{tilte}), (\ref{tildeT2}), and (\ref{deltaT2}),  denoting the resulting sum of the coefficients of $(\rho v_i^2/\tau_i)\tiln^2$  in eqns.
(\ref{tthetav}) - (\ref{tthetaei}) by $\Psi$ and using eqn. (\ref{tei}) we have
\begin{equation}\label{thetat}
\theta T = \Psi\frac{\rho v_i^2}{\tau_i}\tiln^2=0.93\times 10^{-19}n^2T_7^{-1/2}\Psi\tiln^2 \rm{erg}\rm{cm}^{-3}\rm{s}^{-1}.
\end{equation}
The factor $\Psi$ (which is a function of $\omega\tau_i$) is the black curve plotted in Fig. \ref{figure7aab}. 

In following the heating associated with any particular wave propagating outwards from the cluster center, it is important to consider its attenuation. Thus, although entropy production increases outward in the full Braginskii model, this is more than offset by the decreasing amplitude. If ion transport is suppressed, there is very little attenuation and nearly constant entropy production once the wave enters the isothermal regime, resulting in a much flatter heating rate.

\begin{figure}
\begin{center}
\includegraphics[height=55mm]{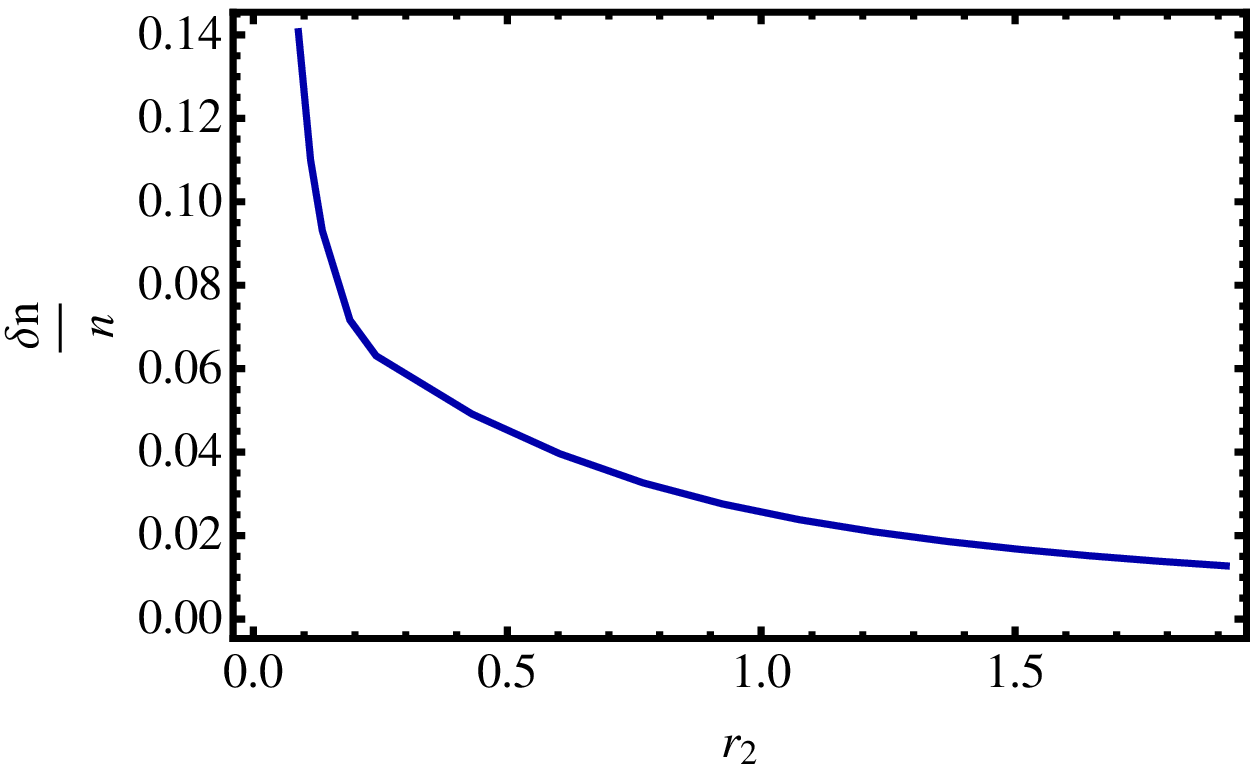}
\includegraphics[height=55mm]{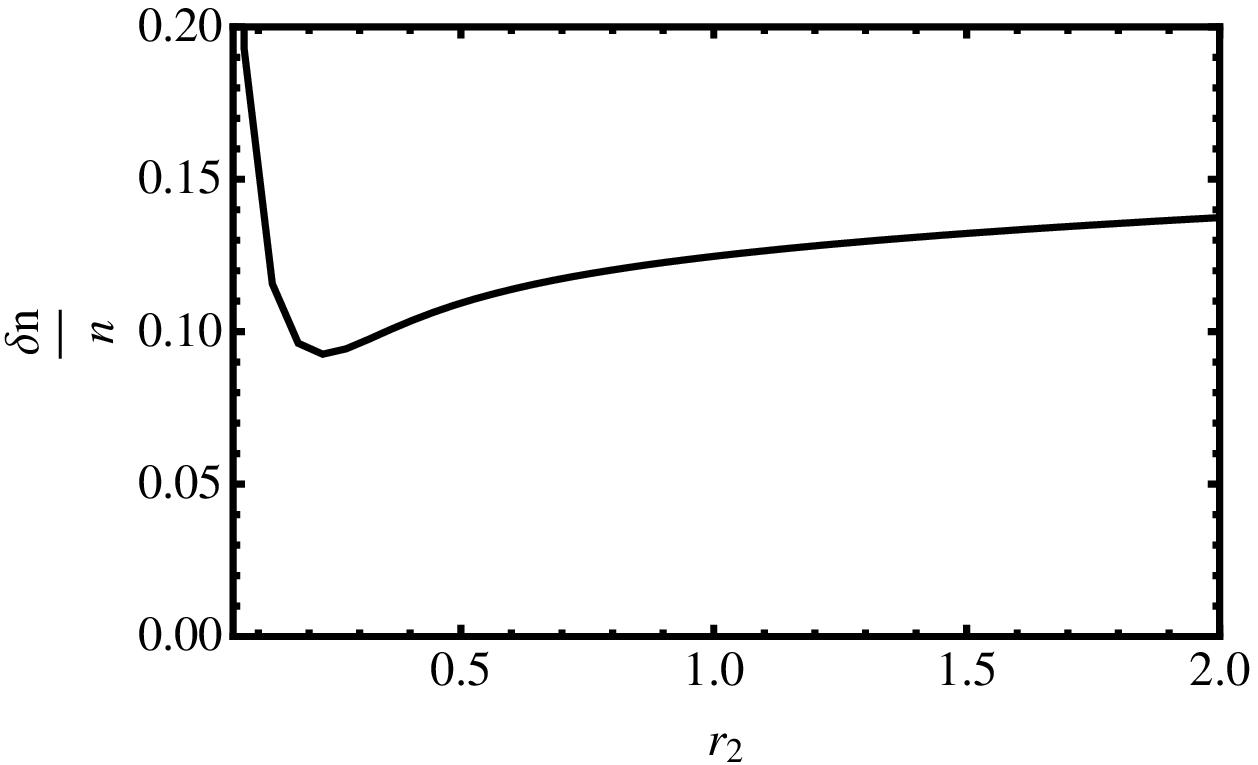}
\caption{The wave amplitude, measured by the relative density perturbation $\tiln=\delta n/n$, at which radiative cooling balances wave dissipation for a wave with a period of
10 Myr in A2199, computed using the results plotted 
in Figure \ref{figure7aab} according to two transport models: full Braginskii (top) and electron thermal conduction only (bottom). Wave amplitudes above the curve would overheat the cluster; lower amplitudes would underheat it. The top plot
underestimates $\delta n/n$ because the viscous damping rates computed from fluid theory are overestimates.}
\label{figure8}
\end{center}
\end{figure}

Equation (\ref{thetat}), because of its $n^2$ dependence, is readily compared to the optically thin radiative cooling rate to determine, for any temperature, $\omega\tau_i$, and choice of $(\xi_{ce},\xi_{ci},\xinu,\xi_{ei})$, the wave relative amplitude $\tiln$ such that wave heating balances radiative cooling. Following F05 we write the radiative cooling rate as
\begin{equation}\label{n2Lambda}
n^2\Lambda= 10^{-24}n^2\left(1.13T_7^{-1.7}+5.3T_7^{0.5}+6.3\right) \rm{erg}\rm{cm}^{-3}\rm{s}^{-1}.
\end{equation}
Combining eqns. (\ref{thetat}) and (\ref{n2Lambda}) gives for $\tiln_{eq}$, the wave amplitude at which wave heating balances radiative cooling
\begin{equation}\label{equilibrium}
\tiln_{eq}=\frac{3.3\times10^{-3}}{\Psi^{0.5}}\left(1.13T_7^{-1.2}+5.3T_7+6.3T_7^{0.5}\right)^{0.5}.
\end{equation}
%
Figure \ref{figure8}
 plots $\tiln_{eq}$ as a function of $r_2$ for a wave period of 10 Myr for the $n$ and $T$ profiles in A2199 and two 
 transport models: full Braginskii (top) and electron thermal conduction only (bottom). While density perturbations as large as 15\% can be tolerated without overheating the cluster center, this value drops below 2\% at 200 kpc for the
 full Braginskii case. This is  due  to the shorter electron and ion collision times at the cluster center, which reduce the transport coefficients
and weaken the damping. However, it is an underestimate because the fluid model overestimates viscous damping relative to the kinetic model. Notably, the relatively flatness of $\Psi$ in the model without ion transport produces a much flatter curve, as shown in the lower panel of  Fig. \ref{figure8}. 

As shown in Fig. \ref{figure6aab}, kinetic effects reduce the damping rate below the predictions of fluid theory. And, because the heating rates due to ion viscosity and ion
thermal conduction scale roughly as $\omega^2$, increasing the wave period by a factor of 3 would increase $\tiln_{eq}$ by almost the same factor. Nevertheless, our work
supports the conclusion reached in F05: in order to balance wave heating and radiative cooling, transport processes must be strongly  suppressed.

Although we have not performed a full stability analysis of acoustic wave heating, Fig \ref{newnoionfig}  and eqn. (\ref{thetat}) suggest that heating by waves in the full Braginskii
model is thermally unstable while heating due to dissipation by electron conduction alone is stable. According to the second equality in eqn. (\ref{thetat}) the heating rate per volume
$H$ for a wave of
fixed amplitude $\delta n\equiv n\tilde n$ is proportional to $\Psi T^{-1/2}$, while $\Psi$
itself is a function of $\omega\tau_i$, which is proportional to $T^{3/2}/n$. If $n$ and $T$ are perturbed isobarically, such that $\Delta n/n = -\Delta T/T$, then $\Delta\Psi/\Psi = 5/2(\Delta T/T)(\Psi^{'}/\Psi)$, where the prime denotes differentiation with respect to argument.
For full Braginskii transport, $\Psi$ increases roughly linearly with $\omega\tau_i$, so
$\Delta\Psi/\Psi\sim 5/2(\Delta T/T)$ and $\Delta H/H=2\Delta T/T$. That is, a positive temperature perturbation increases the heating rate. The same conclusion applies to models wih suppressed Braginskii transport. In contrast, $\Psi$ is nearly
independent of $\omega\tau_i$ for nearly isothermal acoustic waves, so $\Delta H/H=-1/2(\Delta T/T)$: a positive temperature perturbation decreases the heating rate. This argument 
should not replace a full stability analysis, however. Although given in terms of local quantities, the
analysis here implicitly applies to perturbations on lengthscales and timescales that are large enough to average over the acoustic waves that are assumed to be supplying the heat. A complete stability analysis of these larger scale perturbations would include the response of the acoustic waves to slow changes in the density and temperature of the medium in which they propagate \citep{Zweibel1980,Drury1986} as well as perturbations to the radiative
cooling rate due to these larger scale perturbations. Global gradients in the background medium should be included as well. These tasks are well beyond the scope of this paper, but could be a promising direction in the future.
\section{Conclusions}\label{s:conclusion}


In this paper we revisited the pioneering study by \cite{Fabian2005} (F05) of acoustic
wave dissipation and its effect on thermal balance in galaxy clusters. F05 showed that nearly adiabatic acoustic waves which are damped by plasma thermal conduction and viscosity would damp within one wavelength of their source and overheat the cluster gas
if these transport coefficients have the full Braginskii values. Assuming thermal conduction is suppressed entirely and viscosity is reduced to 10\% of its Braginskii value allowed a model in which radiative cooling balanced dissipation of a power law spectrum of waves. 

Our results are consistent with F05. The wave attenutation due to dissipation can be readily evaluated if transport is
so strongly suppressed that the weak damping formula applies (eqn. \ref{weakattenuation}). For example, we found that waves with a 10 Myr period have an $e$-folding length of 50 kpc in
A2199 if transport is reduced to 13\% of its Braginskii value. Although different combinations of suppression coefficients can achieve this, it requires strong suppression of electron thermal conduction, which
accounts for 87\% of the total transport in the full Braginskii model.


Here we have focused on enlarging the toolkit for wave heating studies rather than creating a full heating model based on a spectrum of waves.
Rather than using a quasi-adiabatic approximation, which applies only for strong collisionality ($\omega\tau_e\ll 1$), we derived and solved a dispersion relation (eqn. \ref{fulldr}) which is based on separate energy equations for electron and ion fluids (eqns. \ref{eenergy} and \ref{ienergy}),  each with its own thermal conductivity and coupled through a Coulomb collision based energy exchange term. This allows each particle species to transition from adiabatic to isothermal as collisionality decreases (\S\ref{ss:fluid}) and yields three modes: the acoustic wave, which is the main topic of this paper, and two nearly isobaric modes, one corresponding to relaxation of a thermal pulse (eqn. \ref{conductionwave2}), and the other corresponding to  electron-ion temperature equilibration (eqn. \ref{relaxationwave2}). The isobaric modes are nonpropagating and dissipate their energy within
one wavelength of their source.

Conductive damping of acoustic waves is  self limiting: efficient conduction reduces the temperature contrast across the wave and thus reduces the dissipation associated with heat flow. This effect
is captured by single fluid theory (e.g. \cite{LandauLifshitz1987}), but the
collisionality  at which the transition occurs depends on
electron-ion thermal coupling; Fig. \ref{figure2aab}.  
For the parameter regimes we studied - wave periods of $\sim 10$ Myr in the inner 200 kpc of A2199 - electrons are isothermal over most of the range if the conductivity has the full Braginskii value while ions are nearly adiabatic.  Damping is then due primarily to electron heat conduction near the cluster center, where the gas is most collisional, and to ion thermal conduction and viscosity at lower collisionality and larger radii.

In \S\ref{sss:equilibration} we evaluated the relative difference between the electron and ion temperature fluctuation and found that for moderate $\Omega$, or $\omega\tau_i >\epsilon^{1/2}$, it can become quite large (Fig. \ref{figure4a}). Likewise, the electron temperature fluctuation becomes much smaller relative to the density fluctuation than
predicted from adiabatic theory, and $\delta T_e$ lags $\delta n$ in phase (Fig. \ref{figure5aab}).

The damping rates, as shown in Fig. \ref{figure2aab} are lower than the rates computed in the adiabatic approximation of F05, but are still high (Fig. \ref{figure3aab}). At 100 kpc from the center of  A2199, for example, waves with a 10 Myr period have been attenuated by more than a factor of $e^{8}$ if transport coefficients are at their full Braginskii values (Fig. \ref{new_attenuation2}). Completely suppressing ion transport reduces the attenuation factor to about 50 if electron thermal conduction is at full strength, which is still large. Partially
suppressing electron thermal conduction does result in less attenuation, but the dependence is weak. For example,  decreasing the conductivity by a factor of 3 reduces the attenuation factor by only 20\%.
It is difficult to reconcile these short damping lengths with observations of  roughly uniform density enhancements under the propagating acoustic wave interpretation. It is also unlikely that such strongly damped waves could lie within the inertial range of a turbulent cascade.

In \S\ref{ss:kinetic} we considered kinetic corrections to the fluid picture by comparing the fluid theory to the Fokker-Planck calculation of \cite{KulsrudOno1975} (Figure \ref{figure6aab}). The damping rates calculated according to the fluid and Fokker-Planck models are qualitatively similar, but are higher by $\sim$50\%  in the fluid model under
cluster conditions. In the fully collisionless limit ($k\li\ge\sim 6$), acoustic waves  damp within one wave period unless  the electron temperature is much higher than the ion temperature. Thus, while kinetic effects somewhat mitigate the rapid damping problem under partial collisionality they do not solve it completely and imply rapid damping rates in the  hottest clusters. In fact, without strong suppression of collisionless damping, acoustic waves cannot propagate in the outer parts of galaxy clusters.

In \S\ref{s:heating} we evaluated the heating associated with wave dissipation by calculating the rate at which ion viscosity, electron and ion  heat conduction, and electron - ion
temperature equilibration produce entropy. The relative magnitudes of these entropy sources, shown in Figure (\ref{figure7aab}) track their relative contributions to damping.  Electron heat conduction dominates at the highest collisionalities but ion heat conduction and ion viscosity
dominate as $\omega\tau_i$ increases, similar to their contributions to damping. Entropy production by electron-ion temperature equilibration is always small. By writing the heating rate in a form explicitly proportional to $n^2$ (eqn. \ref{thetat}) we were able to solve for the relative density
perturbation amplitude $\tiln\equiv\delta n/n$ at which the rate of wave  dissipation balances the rate of radiative cooling for a given value of the collisionality parameter $\omega\tau_i$ and ambient temperature $T$ (eqn. \ref{equilibrium}). The result, plotted in the upper panel of Figure \ref{figure8} for a 10 Myr period wave in A2199, shows that the equilibrium wave
 amplitude $\tiln_{eq}$ ranges from about 15\% near the cluster center  to less than 2\% at 200 kpc for the full Braginskii model. However, because the damping rates predicted from the fluid model are higher
than the rates predicted from the Fokker-Planck model,  the values $\tiln_{eq}$ computed here are probably  underestimates. The model with electron thermal conduction damping only can tolerate a significantly larger wave amplitude
without overheating, as seen in the lower panel of Fig. \ref{figure8}.

Bearing in mind that our treatment only applies to plane waves in a uniform medium, and ignores  global geometry, density, and temperature gradients, we can draw some provisional conclusions. Our calculations reinforce the claim of F05 that without significant suppression of transport, acoustic waves in galaxy cluster plasmas should dissipate in 1-2 wavelengths of their source. This is in conflict with the interpretation of regularly spaced multiple density ridges as propagating acoustic waves, poses problems for theories of acoustic turbulence with a large inertial range, and puts strict upper limits on wave amplitudes to avoid overheating. How can these problems be resolved?

Magnetic fields, which are undoubtedly present in galaxy clusters, can reduce transport. A large scale magnetic field perpendicular to the direction of wave propagation almost completely suppresses heat conduction and viscosity, greatly reducing both damping and heating. While this favorable orientation might hold near AGN-driven bubbles due to sweeping up of the magnetized ICM, it is unlikely to be a solution everywhere in the cluster core. A more general way to reduce transport is to increase the effective collisionality of the medium due to magnetic field fluctuations on small scales. This could occur for electrons due to heat conduction instabilities \citep{RobergClark2016} and for ions due to pressure anisotropy instabilities \citep{Kunzetal2011}. As shown in Figs. \ref{newnoionfig}, \ref{new_attenuation1}, \ref{new_attenuation2}, and \ref{figure7aab},
completely suppressing ion transport results in a dissipation rate that is strongly peaked around the location where the electrons transition from adiabatic to isothermal, reduced
attenuation factors, and a relatively flat rate of scaled entropy production with collisionality  and, implicitly, with position in the cluster. The numerical examples presented are for waves with 10 Myr period; the transition occurs closer to or further from the source depending on whether the wave period is shorter or longer. 

It is also possible that the density fluctuations around AGN cavities are driven by a large scale instability which maintains them despite strong dissipation mechanisms. Cosmic ray streaming can destabilize acoustic waves \citep{Drury1986,Begelman1994} but requires  magnetic field strengths and cosmic ray pressures that exceed current estimates for galaxy clusters. Further exploration of these and other instabilities, as well as detailed modeling of acoustic wave propagation and damping for realistic galaxy cluster sources and geometries, are  topics for future work.

\acknowledgements{EGZ is happy to acknowledge support from the NSF through grants PHY 0821899 and AST 1616037 and from the University of Wisconsin through the Vilas Trust and WARF Foundation, as well as the hospitality of the University of Chicago, where part of this work was completed. MR acknowledges NASA ATP 12-ATP12-0017 grant and NSF grant AST 1715140. CSR thanks NASA for support under grant NNX17AG27G. HYKY acknowledges support from NSF grant AST 1713722, NASA ATP (grant number NNX17AK70G) and the Einstein Postdoctoral Fellowship from NASA (grant number PF4-150129). ACF acknowledges ERC Advanced Grant 340442. We thank the anonymous referee for raising interesting questions and pointing out many salient references.